%% file: main.tex
\documentclass[lettersize,journal]{IEEEtran}
\usepackage{amsmath,amsfonts}
\usepackage{algorithmic}
\usepackage{array}
\usepackage[caption=false,font=normalsize,labelfont=sf,textfont=sf]{subfig}
\usepackage{textcomp}
\usepackage{stfloats}
\usepackage{url}
\usepackage{verbatim}
\usepackage{graphicx}
\usepackage{threeparttable}
\usepackage{siunitx}
\usepackage{url}
\usepackage{placeins}
\usepackage{xcolor}

\def\BibTeX{{\rm B\kern-.05em{\sc i\kern-.025em b}\kern-.08em
    T\kern-.1667em\lower.7ex\hbox{E}\kern-.125emX}}
\usepackage{balance}
\begin{document}
\title{Dustin: A 16-Cores Parallel Ultra-Low-Power Cluster with 2b-to-32b Fully Flexible Bit-Precision and Vector Lockstep Execution Mode}
\author{\IEEEauthorblockN{
Gianmarco Ottavi\IEEEauthorrefmark{1},
Angelo Garofalo\IEEEauthorrefmark{1},
Giuseppe Tagliavini\IEEEauthorrefmark{2},
Francesco Conti\IEEEauthorrefmark{1}, \\
Alfio Di Mauro\IEEEauthorrefmark{3},
Luca Benini\IEEEauthorrefmark{1}\IEEEauthorrefmark{3}, and
Davide Rossi\IEEEauthorrefmark{1}}
\thanks{This work was supported in part by EU Horizon 2020 Research and Innovation projects The European Pilot under Grant 101034126, in part by WiPLASH under Grant 863337, in part by ECSEL Horizon 2020 project AI4DI under Grant 826060, and in part by GreenWaves Technologies.}

\IEEEauthorblockA{\IEEEauthorrefmark{1}Department of Electrical, Electronic and Information Engineering (DEI), University of Bologna, Italy}

\IEEEauthorblockA{\IEEEauthorrefmark{2}Department of Computer Science and Engineering (DISI), University of Bologna, Italy}

\IEEEauthorblockA{\IEEEauthorrefmark{3}IIS Integrated Systems Laboratory, ETH Zurich, Switzerland}}


\maketitle

\begin{abstract}
Computationally intensive algorithms such as Deep Neural Networks (DNNs) are becoming killer applications for edge devices. Porting heavily data-parallel algorithms on resource-constrained and battery-powered devices while retaining the flexibility granted by instruction processor-based architectures poses several challenges related to memory footprint, computational throughput, and energy efficiency. Low-bitwidth and mixed-precision arithmetic have been proven to be valid strategies for tackling these problems. We present Dustin, a fully programmable compute cluster integrating 16 RISC-V cores capable of 2- to 32-bit arithmetic and all possible mixed-precision combinations. In addition to a conventional Multiple-Instruction Multiple-Data (MIMD) processing paradigm, Dustin introduces a Vector Lockstep Execution Mode (VLEM) to minimize power consumption in highly data-parallel kernels. In VLEM, a single leader core fetches instructions and broadcasts them to the 15 follower cores. Clock gating Instruction Fetch (IF) stages and private caches of the follower cores leads to 38\% power reduction. The cluster, implemented in 65 nm CMOS technology, achieves a peak performance of 58 GOPS and a peak efficiency of 1.15 TOPS/W.
\end{abstract}


\section{Introduction}

Modern Near-Sensor Analytics Applications (NSAA) increasingly require running complex workloads such as Deep Neural Networks (DNN) on Internet of Things (IoT) end-nodes. These devices are severely constrained in terms of power envelope, memory, and cost (i.e., silicon area and technology). An emerging trend to tackle this problem is to employ the most compact data representation usable from a numerical viewpoint for each given task of an application, exploiting low-precision and mixed-precision arithmetic operations, reducing the complexity of arithmetic units and the memory footprint of an application~\cite{van2020bayesian,cai2020rethinking}. 

From an architectural viewpoint, extreme low-bitwidth mixed-precision arithmetic has been mainly applied in specialized accelerators~\cite{GANPU, lee2018unpu}. Exploiting this technique with dedicated hardware is very effective since the whole datapath is typically designed for a single or a subset of functions, safely tuning each operation to the desired precision. However, applying this principle to fully programmable architectures is challenging since the algorithms to be executed are not known a priori. Consequently, multiple formats must be supported, increasing the complexity and overheads of instruction fetch (IF) and decode (ID) stages.

Still, previous work on low-bitwidth computations on instruction processors demonstrated promising results, especially in DNN inference. Garofalo~\textit{et al.}~\cite{garofalo2020pulp} proposed a C library for DNN inference exploiting 8-bit Single Instruction Multiple Data (SIMD) instructions, as well as other Digital Signal Processing (DSP) Instruction Set Architecture (ISA) extensions. This solution outperforms a commercial library (CMSIS-NN~\cite{lai2018cmsis}), implementing the same functions on ARM Cortex M4 and M7 by 4.54$\times$ or 2.54$\times$ respectively, only featuring support for 16-bit SIMD instructions. While 8-bit SIMD is now widely supported by all the major ISAs, such as ARMv8.1 helium \cite{ARMHELIUM}, more aggressive approaches have been presented in Garofalo~\textit{et al.}~\cite{garofalo2020xpulpnn}, where SIMD support has been extended to 4-bit and 2-bit operations, leading to further performance and energy efficiency gains.

However, the extensions proposed in Garofalo~\textit{et al.}~\cite{garofalo2020xpulpnn} only tackle part of the challenge, lacking featuring support for mixed-precision operations (i.e., operations where the source operands have different, reduced bitwidths, e.g., 2-bit and 4-bit). Mixed-precision execution requires data conversion and packing/unpacking operations leading to significant overheads if not natively supported by the underlying hardware~\cite{bruschi2020enabling}. When applied to DNNs, exploiting mixed-precision computations on state-of-the-art processors dramatically reduces the memory footprint enabling the execution of MobileNets on tiny end-nodes. However, it comes with a significant performance overhead over uniform SIMD. Furthermore, supporting too many mixed-precision formats leads to a proliferation of instructions. For example, supporting SIMD instructions for efficient execution of DNNs leads to more than 300 instructions due to precision format ranging from 16- to 2-bit and all possible permutations. This effect increases the complexity of the Instruction Fetch (IF) and Instruction Decode (ID) stages and possibly saturates the ISA encoding space. Finally, when moving from single to multi-core architecture, the cost of fetching and decoding instructions increases linearly with the number of cores, even if all the cores are executing the same instructions on different data (e.g., computing different elements of an output activation tensor in a DNN).

We present Dustin, a low-power IoT processor with a software-programmable accelerator composed of 16 RISC-V cores optimized for energy-efficient end flexible execution of integer mixed-precision operations to tackle the presented challenges. The DSP cores support mixed-precision extensions through 2b-to-16b SIMD instructions, accelerating arithmetic operation and complex packing-unpacking and conversion operations required by mixed-precision computations. The format of input operands is set through a dedicated control register to reduce the complexity of the IF and ID stages of the processors. 

Furthermore, the cluster can be dynamically configured into a fine-grain \textit{Vector Lockstep Execution Mode} (VLEM), turning off the IF stages and private instruction caches of all the cores except one. This technique boosts the energy efficiency of data-parallel sections of the code, reducing power consumption by 38\% on average with no performance degradation on critical data-parallel kernels while still offering Multiple Instructions Multiple Data (MIMD) flexibility for general-purpose code.

The contributions of the paper are the following:
\begin{itemize}
\item{A 16-cores cluster implemented in 65nm CMOS technology featuring 16 mixed-precision cores supporting fully-flexible 2b-to-32b bit precision scalability.}
\item{A dynamically configurable vector lockstep execution mode allowing to switch-off IF stages and private instruction caches of slave cores, saving up to 38\% of cluster's power consumption.}
\item{The full software stack for the proposed SoC, including compiler support as well as a programming model for efficient exploitation of two architectural features described above.}
\item{The evaluation of the proposed architecture on the full inference of a ResNet8 DNN quantized with 8-bit activations and 4-bit weights, demonstrating 2.7$\times$ performance boost and 4.2$\times$ energy efficiency boost when enabling the two key features of the cluster.}
\end{itemize}

Implemented in robust and cost-effective 65 nm CMOS technology, Dustin achieves 15 GOPS and 303 GOPS/W on 8-bit integer arithmetic. These results are comparable to SoA fully programmable systems implemented in much more scaled technology nodes (40 nm and 22 nm) -- with a further boost in performance (3.7$\times$) and efficiency ($1.9\times$) on low-bitwidth mixed-precision workloads, up to 58 GOPS and 1.15 TOPS/W nearing the efficiency of dedicate accelerators.

\section{Related Work}
The deployment of DNN into tiny devices has been facilitated by quantization and mixed-precision computing. This section presents a review of the state of the art of quantization techniques and recent architectures for QNN inference.

\subsection{Quantization}
Quantized inference and training of DNNs are both computationally intensive. In this context, an efficient representation of numerical values is particularly important since state of the art DNN models are heavily over-parameterized, providing ample opportunity for reducing bit precision without impacting accuracy~\cite{howard2017mobilenets, sandler2018mobilenetv2}.

The use of lower precision quantization improves hardware performance but can lead to significant accuracy degradation. Mixed-precision quantization, in which each layer
is quantized with a different bit precision, addresses this issue but presents a challenge in selecting the appropriate bit setting for each layer. Different methods have been proposed to address the large search space for mixed-precision
quantization, including reinforcement learning~\cite{wang2019haq}, Neural Architecture Search~\cite{wu2018mixed}, and regularization-based approaches~\cite{naumov2018periodic}. HAWQ~\cite{dong2019hawq} is an automatic method that uses second-order sensitivity to find mixed-precision settings, which has been shown to be faster than other methods. HAWQv2 and HAWQv3~\cite{dong2020hawq, yao2021hawq} have been proposed to improve this method with integer-only and hardware-aware quantization, respectively, and have been shown to be efficient on T4 GPUs, with up to 50\% speedup compared to INT8 quantization.

Extreme quantized neural networks are another trend in recent deep neural networks design for embedded computing. For example, Choi et. al.~\cite{choi2018bridging} demonstrate the use of a 2-bit CNN using a uniform quantization method, where different techniques are employed for quantizing activations and weights. The PACT technique, which finds an optimal value for the clipping threshold of the RELU function, is used for inputs during training, and SAWB, a scheme that aims to minimize quantization error without extensive search, is used for weights. The authors show that a 2-bit quantization can result in 3\% accuracy degradation on the ResNet18 and ResNet50 model for Imagenet datasets compared to a full precision network (8-bit).

\subsection{Dedicated Accelerators for DNNs}

On the Hardware side, we have several approaches to the end-to-end inference of a DNN. They range from dedicated accelerators to software programmable solutions more similar to those proposed in this work. We can categorize them by performance, flexibility, and power envelope. The first category includes specialized accelerators that trade off flexibility to excel in performance and efficiency. A common feature of all the presented solutions is the exploitation of low bit-width (down to byte or even sub-byte) to improve the efficiency of inference. We detail a few notable approaches below.

Envision~\cite{moons201714} is a DNN accelerator with a reconfigurable computational engine capable of using bit precision of 1-16b. It employs a circuit-level voltage and frequency scaling technique to improve efficiency, capable of peak 76~GOPS and averages 2~TOPS/W. Thinker~\cite{yin20171} employs configurable 2D arrays that can be partitioned into sub-arrays to compute different types of layers. Each PE presents a set of two 8-bit multipliers that can be merged for a 16-bit operation or can compute two 8-bit (or less) in one cycle. It peaks at 380~GOPS and 5~TOPS. Loom \cite{sharify2018loom} is a bit-parallel DNN accelerator where performance and efficiency scale inversely with weight and activations precision. Loom can also reduce precision dynamically by inspecting groups of 256 activations that it processes concurrently, further increasing the effectiveness of bit reduction on the overall efficiency. EyerissV2~\cite{chen2019eyeriss} is a DNN accelerator that connects multiple PE clusters with a flexible NoC. The network can be configured to efficiently work on either high-bandwidth for networks that present low-reuse or, when reuse is high, it can still exploit spatial data reuse (via multicast or broadcast) to achieve high energy efficiency. It uses a fixed weight/activation precision format of 8-bit and is capable of a peak throughput of 153.6 GOPS.

\subsection{FPGA based accelerators}

DNN accelerators can also be deployed into FPGAs. This solution provides extra flexibility compared to ASICs since they can be reconfigured with Hardware Description Language (HDL), but this comes at an order of magnitude less performance and efficiency. The power envelope raises to Watt level for these devices, which can be problematic for battery-powered devices.
A new family of FPGAs announced by Lattice, namely Sense-AI~\cite{LatticeSENSEAI}, provides comprehensive hardware and software solutions for always-on artificial intelligence within a power budget between 1 mW and 1 W. Despite that, these ultra-low-power FPGA families have limited LUT capabilities and are still too expensive for many applications where MCUs are traditionally chosen for their low cost.
The adoption of FPGAs remains an obstacle for the average IoT programmer, who demands the highest flexibility from microcontroller systems. This work focus on more flexible and user-friendly solutions based on software programmable instruction processors.

\subsection{Software Programmable Solutions}

\subsubsection{Low-Power MCUs}

Given that the rise of DNN and quantization is relatively recent, "classical" commercial microcontroller (MCU) cores such as Cortex M4 and M7 struggle to compete with newer architectures. This is shown in \cite{garofalo2020pulp} where a RISC-V core\footnote{https://github.com/openhwgroup/cv32e40p} significantly outperforms the ARM counterpart in CNN layers with 3.2$\times$ to 6$\times$ when using the de-facto standard quantization of 8-bit.

To address the DNN computing at the edge, ARM presented the Cortex M-55 core based on the ARMv8-1M ISA. The core's general-purpose performance sits between an M4 and M7 with an M-Profile Vector Extension (MVE) called Helium supporting 8-bit MAC instructions. The vector extension uses a 64-bit data interface, meaning it can execute 2$\times$32-bit, 4$\times$16-bit, 8$\times$8-bit fixed-point operations per cycle \cite{ARMHELIUM}.

Specialized accelerators can also be found in the microcontroller class of devices side by side with general-purpose cores. In~\cite{VEGA}, a cluster of 9 RISC-V cores with a tightly-coupled CNN accelerator improves performance by 3.3$\times$ and double efficiency w.r.t. executing the same network on the software-programmable cores. ARM adopts the same approach coupling the Cortex M-55 with the optional Ethos-55, an accelerator designed to boost machine learning tasks; depending on the configuration, the system can execute 32 to 256 MAC/Cycles. This solution can help mitigate the effect of under-utilization on ASIC acceleration, mixing high-throughput (when possible) with high flexibility. 

In~\cite{garofalo2020xpulpnn} a RISC-V ISA extension called Xpulpnn is proposed. It expands on the already available DSP instructions to support the sub-byte precision format of 4- and 2-bit. It introduces \textit{MAC\&LOAD} instructions that simultaneously execute the dot-product while loading an operand for the next operation. Xpulpnn outperforms the commercially available M4 and M7 from 2.8$\times$ to 19.2$\times$ on Quantized convolutional layers. When going mixed-precision, the efficiency boost of Xpulpnn w.r.t. ARM cores narrows significantly because of the massive software overhead necessary for packing and unpacking data. Dustin's cores have direct hardware support for mixed-precision operations, eliminating performance degradation compared to uniform precision. To the best of the author's knowledge, no microcontroller class device supports mixed-precision instructions with dedicated ISA extensions.

\subsubsection{SIMD vs MIMD}

General-purpose multi-core CPUs offer a great deal of flexibility but fall behind in efficiency when exploiting data-level parallelism compared to SIMD-style architectures. General-purpose multi-core architectures require per-core hardware for fetching and decoding instructions creating significant hardware and power overhead, a phenomenon also known as the ”Von Neumann bottleneck” (VNB).

To Mitigate the VNB, GPUs can efficiently capitalize on data-level parallelism, given that every multi-core multi-threaded cluster needs only one unit that fetches and dispatches instructions to multiple execution units. A problem that arises when executing code in GPUs is branch divergence: a piece of code can contain branches. Different data values determine different branch outcomes: some of the threads take the if-path while others do not, causing the code to be executed sequentially and degrading performance. Specifically, on NVIDIA GPUs, this was dealt with by thread masking, where a mask with one entry per thread would indicate whether to execute the branch or not. In the proposed architecture, this situation can be avoided by just switching from VLEM to MIMD mode.

In~\cite{dogan2013synchronizing} Dogan~\textit{et al.} propose a system of 8 general-purpose cores that can synchronize and execute instructions in lockstep. On branch divergences, the cores get out-of-sync, execute the code simultaneously, and wait on a convergent point on a synchronization barrier for resuming lockstep execution. In conjunction with the lockstep, a broadcast mechanism serves multiple same-address memory requests as one, allowing significant power savings. However, the proposed architecture has 2 significant shortcomings: 1) Power management is not applied to the instruction fetch units of the slave cores, and 2) The instruction memory hierarchy of the core is based on simple scratchpad memories, which is somehow not realistic for a high-performance IoT processor.

In our work, the cluster can be configured dynamically to work in a MIMD or Vector Lockstep Execution Mode (VLEM). In VLEM, only one core fetches instructions (similar to a GPU) and dispatches them to the remaining 15 cores. This design allows to clock-gate the private I\$ and Instruction fetch stages of the remaining cores enabling substantial power savings (in addition to the broadcasting feature). On divergent branches, the cluster is configured back to MIMD mode allowing simultaneous execution of each branch. This approach gives higher flexibility when compared to GPUs and, at the same time, removes the power overheads inherent in general-purpose CPUs that execute SIMD code.


\section{SoC Architecture}
\begin{figure}[t]
    \centering
    \includegraphics[width=0.95\linewidth]{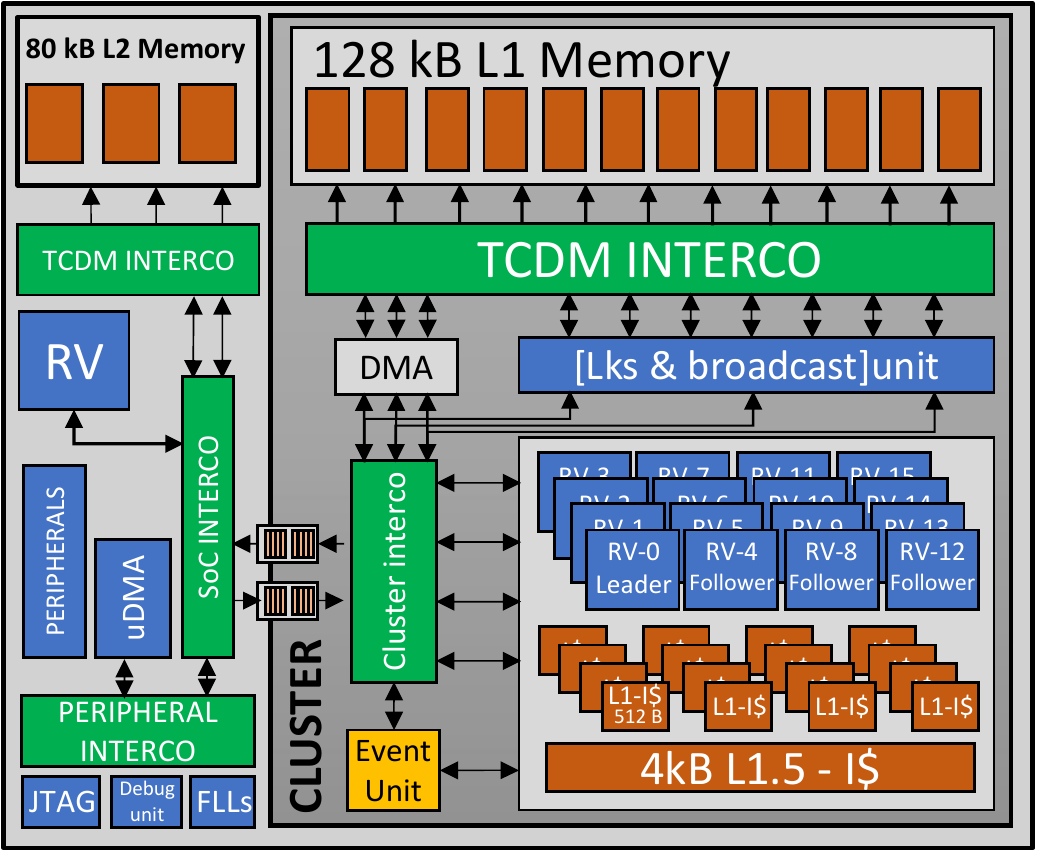}
    \caption{Overview of the Dustin SoC Architecture.}
    \label{fig:soc_archi}
    \vspace{-0.5cm}
\end{figure}

Fig. \ref{fig:soc_archi} shows the architecture of Dustin SoC. The main contribution of the work refers to the RISC-V cores compute cluster. The IPs surrounding such domain, i.e., the RISC-V core named Fabric Controller (FC), a standard set of peripherals, and an L2 memory storing the code executed by both the compute cluster and the FC, as well as the FLLs for clock generation, serve as a programmable testbench.

The programmable cluster resides in a dedicated power and clock domain, communicating with the SoC subsystem with two AXI 4 ports (one initiator and one target) connected to the SoC interconnect through dual-clock first-in-first-out buffers (FIFOs).
The cluster is built around 16 32-bit RISC-V processors; the FC turns on and sets the frequency of these cores when it offloads the computation to the programmable accelerator.

The cores share a 128 kB L1 memory interleaved on 32 banks. The number of cores and banks chosen for the implementation of the proposed cluster can be considered an upper bound for tightly-coupled clusters of processors featuring a single-cycle latency interconnect (LIC), as increasing the number of cores would impact timing, area, and power significantly. Further performance scaling might be achieved by introducing pipelined interconnects, such as in Cavalcante et. al.~\cite{cavalcante2021mempool}, or instantiating multiple clusters in the design, such as in Benini et. al.~\cite{benini2012p2012}.
The \textit{Lockstep} unit and the \textit{Broadcast Unit}, serving the requirements of VLEM are interposed between the cores and the LIC to enable a reconfigurable SIMD/MIMD execution model of the cluster. This key innovation will be discussed in more detail in Section~\ref{sec:VLEM}.

The cluster can access peripherals such as the timer, event unit, DMA, and AXI4-bus via a dedicated peripheral interconnect. The DMA manages data transfer between the L2 and L1 memory, featuring 2-D data transfers and up to 16 outstanding transactions to hide the latency between the two levels of the memory hierarchy.
The cores share a 2-level latch-based instruction cache. The latch-based design allows to save up to 4$\times$ on instruction memory reads \cite{meinerzhagen2010towards}, while the two-level nature of cache reduces long critical paths in the instruction fetch stage while improving effective cache capacity. The first level (512 B) is private, and the second level (L1.5) is a 4 kB 8-banks shared cache connected to the L1s with an interconnect similar to the LIC with low latency. The L1.5 refills from the L2 memory hosting resident code.

For efficient parallel computing, the cluster supports parallel thread dispatching and synchronization via a dedicated hardware block, the \emph{event unit} \cite{glaser2020energy}. The cores can wait on events by doing loads on aliased, low-latency memory-mapped registers of the event unit. In addition, the event unit controls the clock-gating for all the cluster cores, meaning that a core waiting for an event can be put to sleep immediately and can resume after an event in two cycles.

\subsection{Bit-Scalable Precision Processor}
\begin{figure}[t]
\centerline{\includegraphics[width=0.5\textwidth]{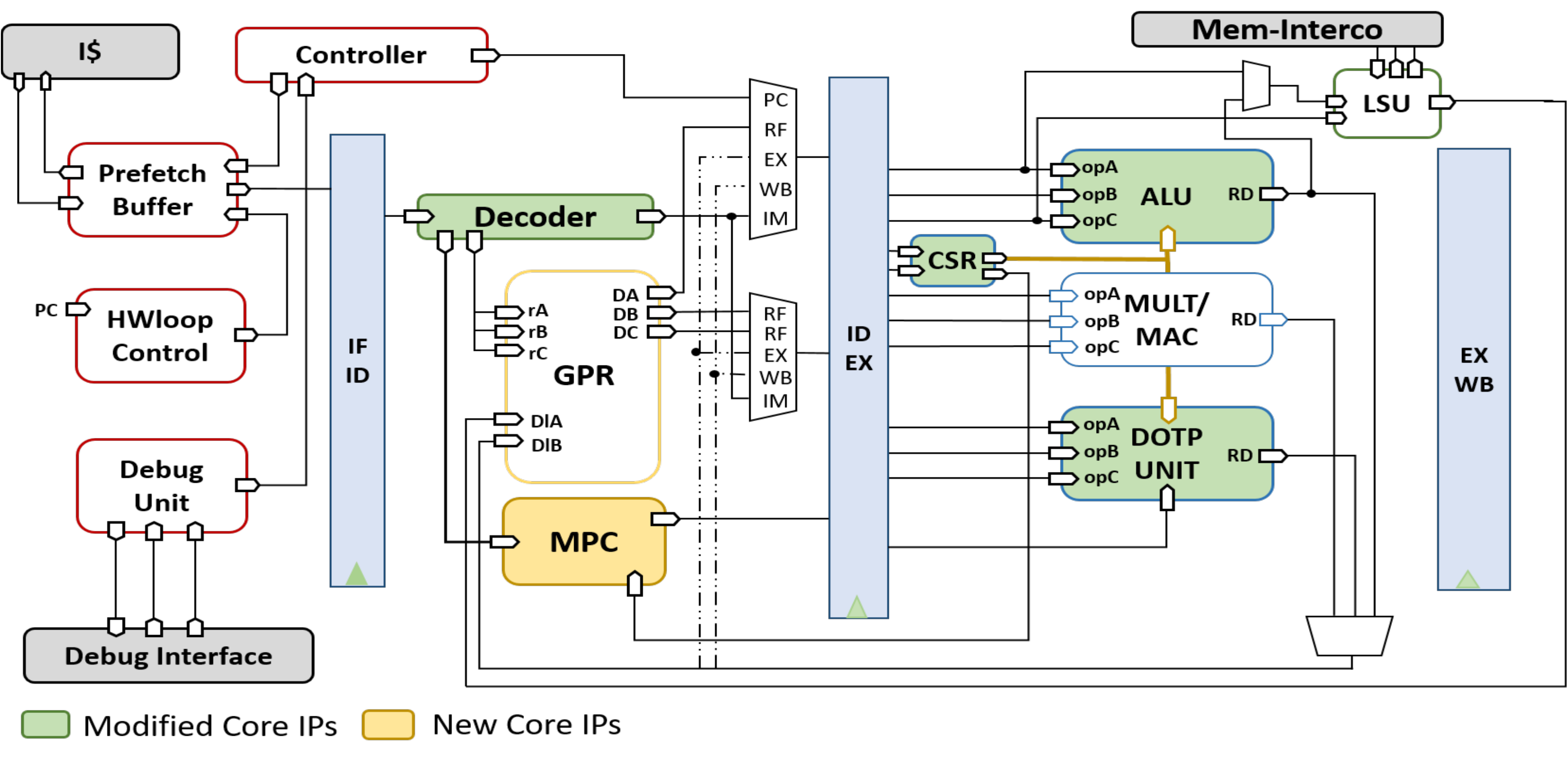}}
\caption{Dustin's Cores, an extension of CV32E40P.}
\label{fig:mpic}
\vspace{-0.5cm}
\end{figure}

The cores employed in Dustin extend RI5CY \cite{garofalo2020pulp}, a 4-stages pipeline in-order single-issue processor. Fig.~\ref{fig:mpic} shows a diagram of the cores' pipeline: changed submodules w.r.t. the baseline RI5CY are highlighted in green, whereas the entirely new blocks are shown in yellow. The baseline RI5CY supports the standard RISC-V extensions (I, M, C, and F) and implements a domain-specific extension, called \textit{XpulpV2}, that introduces several features useful to improve DNN inference such as hardware loops, bit manipulation instructions, load/store with post-modified access, SIMD operations for 16- and 8-bit format \cite{garofalo2020pulp}. 

The key novel efficiency-boosting enhancements of the proposed core are: a new mixed-precision SIMD dot product execution unit, integrated into the micro-architecture of RI5CY, providing support for power-of-two precision formats ranging from 16- down to 2-bit and all their possible mixes; an extension of the RISC-V ISA with a set of instructions to deal with mixed-precision SIMD operations through a dynamic bit-scalable execution model. Given the ten precision combinations that the SIMD Unit can execute, using a standard ISA extension approach where we encode one instruction per each type and format of operation would lead to an enormous proliferation in the number of instructions and increased complexity of the decode stage of the micro-architecture. Looking just at the new dot-product instructions, which, however, are not the only supported operations, the standard approach would increase the total amount of instructions from 24 (of the baseline \textit{XpulpV2}) to 120.

\begin{figure}[t]
\centerline{\includegraphics[width=0.5\textwidth]{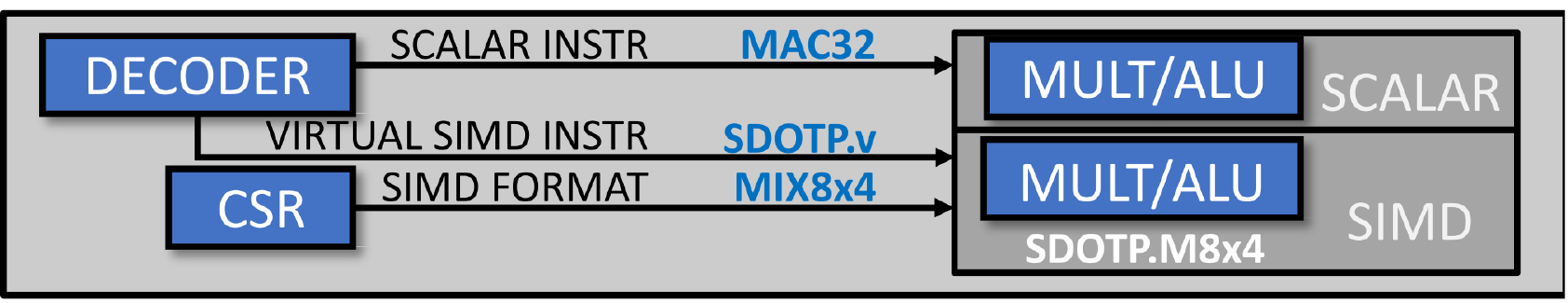}}
\caption{Control signals for SIMD and Scalar instructions. The SIMD instruction is a Sum-of-Dot-Product, and the format is a mixed-precision 8x4. The bottom picture contains the encoding of the formats that are contained inside the CSR.}
\label{fig:sb-dec}
\end{figure}

The dynamic bit-scalable approach exploits a concept we call \emph{virtual instructions}. From users' perspective, the new SIMD instruction work as regular ones; however, the key difference lies in the operation precision, which is not directly encoded into the instruction itself. The precision is specified at run-time by the content of a Control and Status Register (CSR), written in advance by the programmer to set the desired format of the operands. This approach reduces the amount of \textit{dot-product} instructions that need to be encoded in the ISA by 10$\times$.
Fig.~\ref{fig:sb-dec} shows a graphical illustration of the dynamic bit-scalable execution model. The scalar instructions encode the format and type of the operations meaning that the decoder alone can provide full information to forward to the ex-stage. For SIMD instructions, the decoder forwards only information on the type of operations to perform to the ex-stage, i.e., it issues the virtual instruction; the additional control signals required by the execution units to determine the format of the operands are provided by the CSR.
%

\begin{figure}[t]
\centerline{\includegraphics[width=0.5\textwidth]{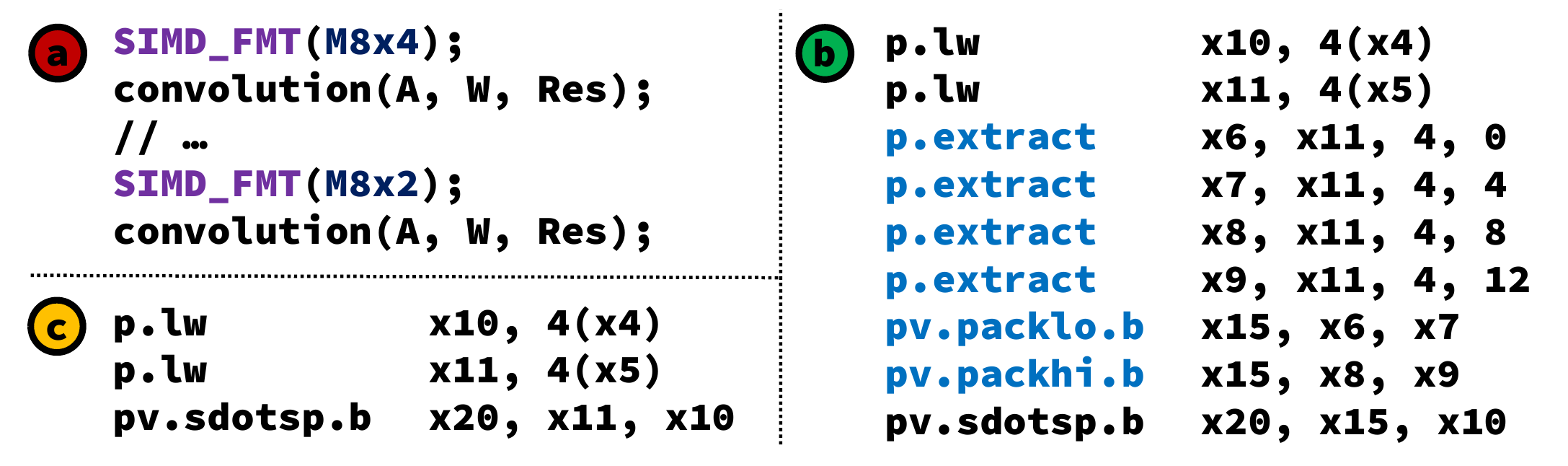}}
\caption{a) Procedure to set precision formats before calling the convolution function. b) Mixed-precision convolution inner loop with data unpacking and conversion overhead. c) Mixed-precision convolution inner loop with direct mixed-precision support.}
\label{fig:coding}
\vspace{-0.5cm}
\end{figure}

In mixed-precision convolution routines, the format of the dot-products can change between layers (or tensors). As illustrated in Fig.~\ref{fig:coding} (a), the programmer can set the CSR to the desired format using the \textit{SIMD\_FMT} macro before calling the kernel. The overhead of this operation is minimal, as it occurs within a single cycle and thus it is negligible compared to the thousands to millions of cycles required for convolution layers and tensor operations.

Fig.~\ref{fig:coding} (b) and (c) highlight the benefits of the mixed-precision support at the ISA level introduced in this work. They compare a snippet of the assembly code of the innermost loop of a mixed-precision 8-bit$\times$4-bit convolution kernel, Fig.~\ref{fig:coding} (b) targeting the \textit{XpulpV2} ISA and Fig.~\ref{fig:coding} (c) the ISA extensions presented in this work. In Fig.~\ref{fig:coding} (b), once we load the 4-bit weights, additional unpacking and packing instructions are inserted to cast the 4-bit SIMD vector to 8-bit before sending the data to the 8-bit DOTP unit. This is achieved with two step: i) \textit{p.extract} instruction takes the 4-bit elements loaded into register x11 and sign-extend them to a 32-bit integer; ii) \textit{pv.packlo/hi.b} takes the first 8-bit of two source registers and pack them in the lower/upper 16-bit of the destination register, x15 in this case. The figure only show partial code of the inner-loop, given that x11 contains 8 elements and x10 only 4, a new load for the activation (x10) in necessary before loading new weights. The unpacking procedure is then repeated for to the upper part of the register x11 (from bit 16 to 32) before the MAC. Nonetheless, this procedure adds 6 instructions of overhead for each dot-product. Thanks to the hardware support for mixed-precision SIMD operations, the execution depicted in Fig.~\ref{fig:coding} (c) requires no additional packing/unpacking instructions, providing a significant boost in performance on Mixed-Precision executions.

At the micro-architectural level, we extend the ALU and the DOTP units of RI5CY to support the ISA instructions introduced above. We add extra CSR registers for storing the operations' formats, and we design a mixed-precision controller to handle the selection, slicing, and routing of SIMD vector elements to the execution units of the \textit{ex\_ stage} of the pipeline.

We detail the DOTP unit architecture, omitting that of the ALU for the sake of conciseness, as its design follows a similar approach. The DOTP unit computes the dot-product (\textit{dotp}) operation between two SIMD registers and accumulates the partial results over a 32-bit scalar register through an adder tree in one clock cycle of latency. The SIMD vectors can be either symmetric or featuring mixed formats, within a precision range from 16-bit down to 2-bit.

\begin{figure*}[t]
\centerline{\includegraphics[width=0.95\textwidth]{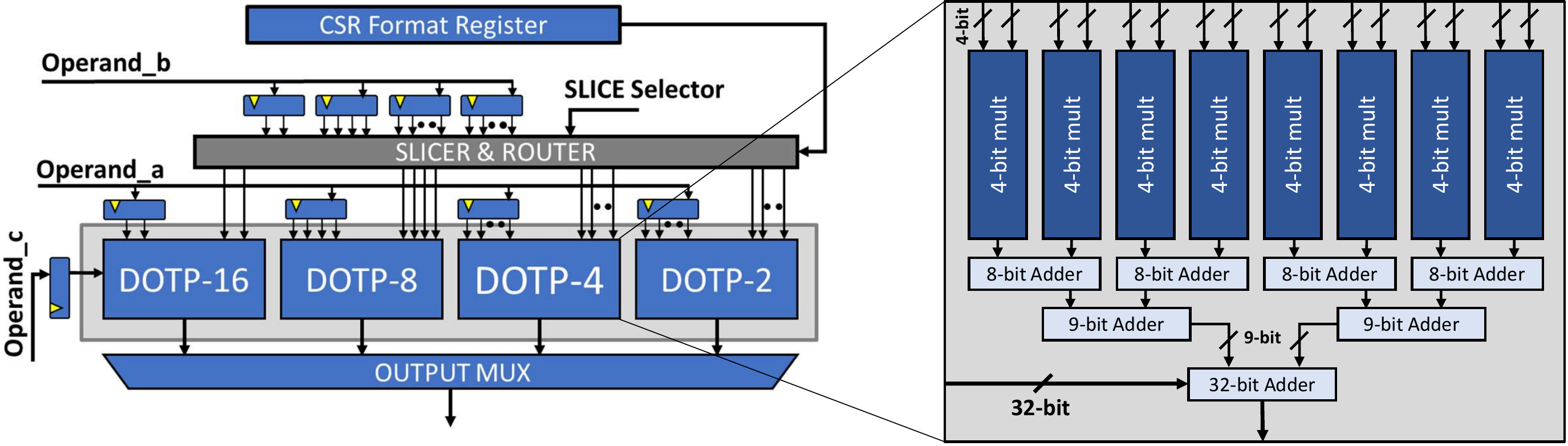}}
\caption{On the left, an overview of the dot-product unit; on the right, the internals of the DOTP-4.}
\label{fig:dotp_unit}
\end{figure*}
\begin{figure*}[t]
\centerline{\includegraphics[width=0.95\textwidth]{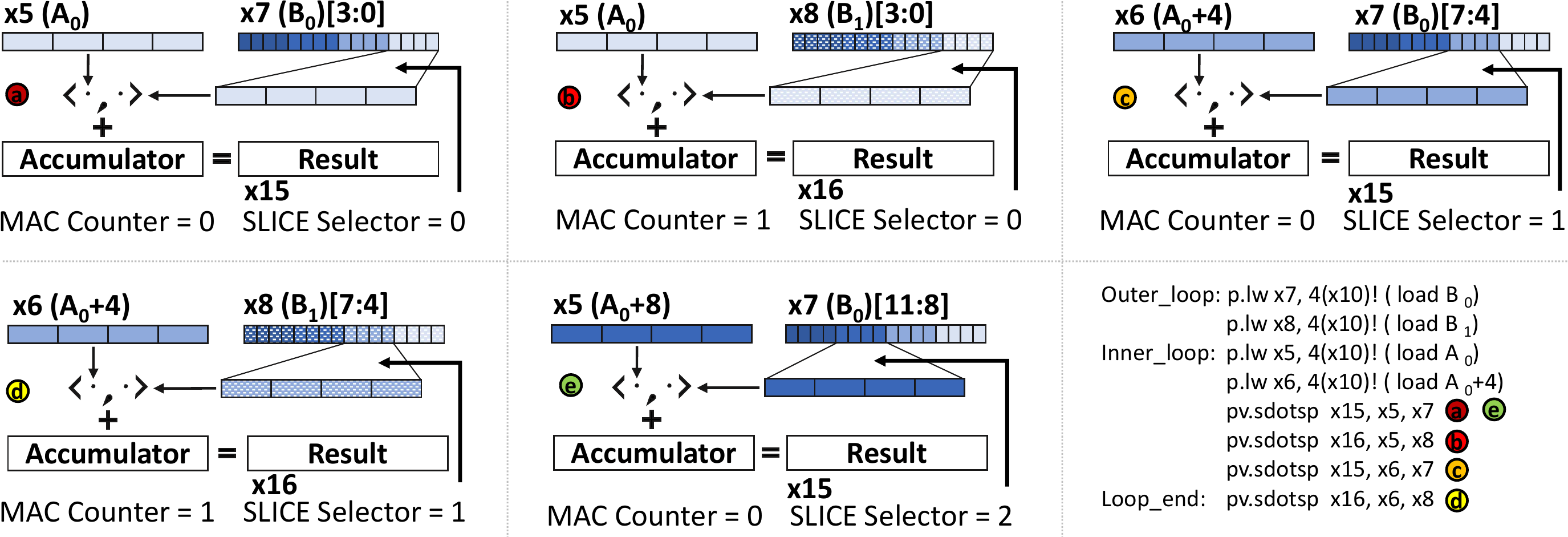}}
\caption{Mixed-Precision Dot Product 8x2, on the bottom right, an example of an inner-loop of a convolution kernel run on Dustin. p.lw is a post-increment load, automatically updating the $A_0$ pointer by 4. Indexes within square brackets indicate the elements of the $B_x$ operand selected by the mixed-precision controller.}
\label{fig:mixed_pic}
\vspace{-0.5cm}
\end{figure*}

The dot-product operations are implemented in the DOTP unit with a number of multipliers equal to the number of elements of the SIMD vector (in the mixed-precision context, the highest-precision SIMD vector determines the correct set of multipliers to be used) defining four different ``bitwidth'' regions, each followed by a dedicated adder tree that sums up the partial products, as shown in Fig.~\ref{fig:dotp_unit} for the 4-bit precision operation (DOTP-4). The sum-of-dot-product (\textit{sdotp}), which is the SIMD equivalent of a MAC operation, is supported by adding an additional 32-bit scalar operand at the input of each adder tree.

To minimize the logic, operand B is designated to be always the smallest operand in mixed-precision SIMD operations without loss of flexibility. Referring to Fig.~\ref{fig:dotp_unit} we introduce a \textit{slicer \& router} network to: a) slice \textit{operand\_b} according to the value coming from \textit{CSR Format Register}; b) select the correct subset of \textit{operand\_b} with the \textit{SLICE selector} signal coming from the mixed-precision controller; c) sign-extend (or zero extend) the vector to match the size of \textit{operand\_a} in order to use the appropriate DOTP unit (e.g., an 8x4-bit operation requires DOTP-8).

Since \textit{dotp} operations are critical from a timing closure viewpoint, replicating the hardware resources over different bitwidth ``regions'' of the DOTP unit avoids impacting the critical path of the RI5CY core at the cost of additional area and power. However, to mitigate the effects of hardware replication on the dynamic power consumption of the system, all the input operands of the DOTP unit are register-gated to avoid switching for operands not involved in the current SIMD operation.

The last component added to the core is the mixed-precision controller, which selects the proper sub-group of elements from the second source register (operand \emph{B}). The entire process is described in Fig~\ref{fig:mixed_pic}. When the decoder identifies a mixed-precision MAC operation, the \emph{slice selector} selects the proper sub-group of \emph{B} operands (e.g., [3:0] in (a)). The group of \emph{B} operands is kept the same until the \emph{MAC counter} reaches its target, defined by a control register explicitly written by the programmer. Once the \emph{MAC counter} reaches the programmed target, it goes back to index 0, while the slice selector is shifted left (e.g., [7:4] in (c)). This process continues until the whole kernel has been executed.

QNN kernels follow a uniform pattern in their computation, and the combination of sequential and reuse information is enough to deal with mixed-precision computation. To give maximum flexibility to the programmer, we implement the ability to control via software the subgroup of operands to use by directly writing the value in the counter. This feature is useful if the application includes an operation pattern that the mixed-precision controller cannot deal with automatically.

\subsection{Vector Lockstep Execution Mode}
\label{sec:VLEM}
\begin{figure}[t]
\centerline{\includegraphics[width=0.5\textwidth]{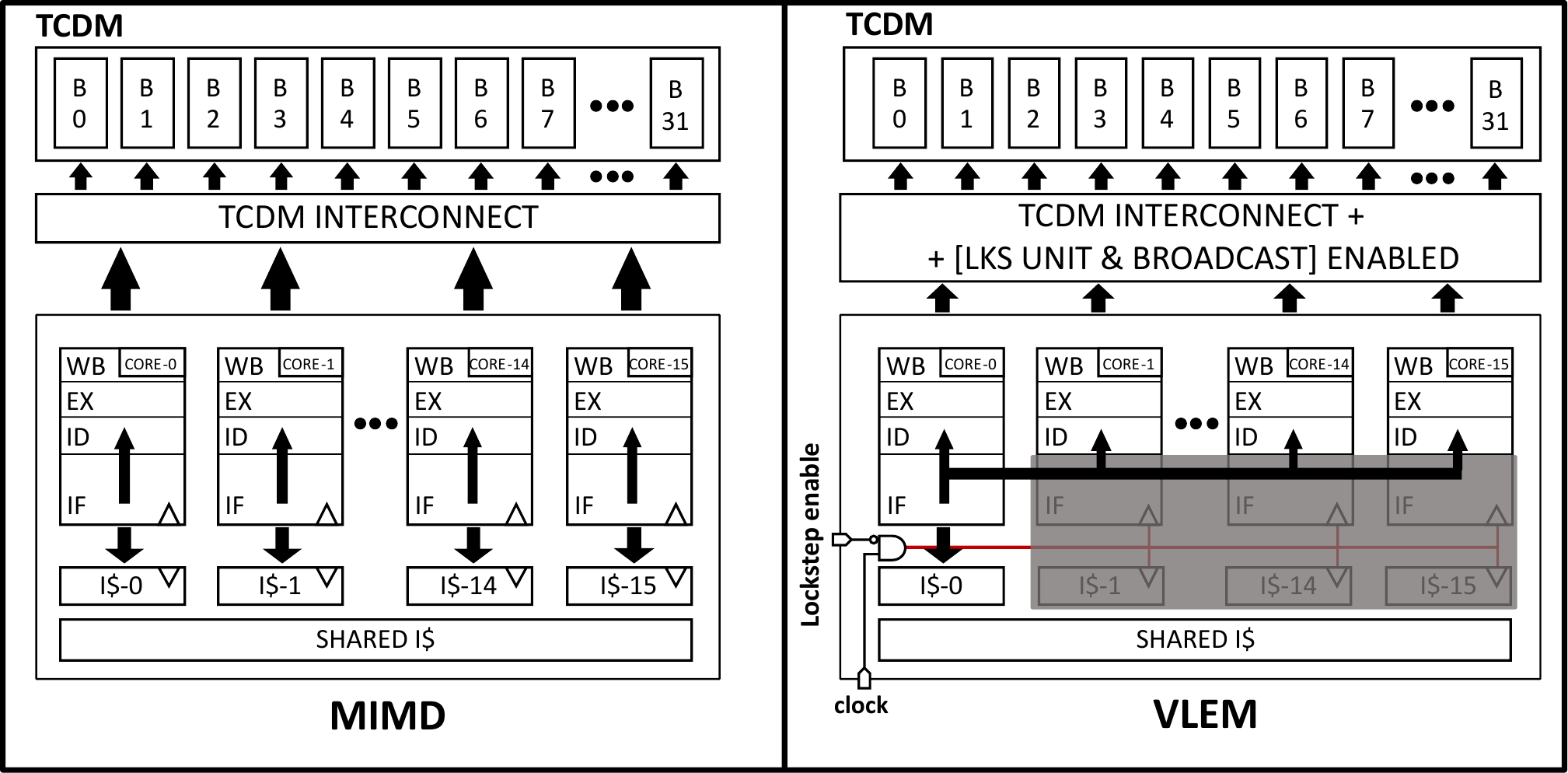}}
\caption{Cluster's diagram for MIMD and VLEM execution models.}
\label{fig:vlem_pic}
\vspace{-0.5cm}
\end{figure}

The inner kernel of intensive workloads can be parallelized on multiple cores that execute the same instruction on different data. On a cluster of general-purpose cores using a MIMD execution model, this translates into a loss of efficiency due to the VNB, which implies extra energy for fetching the same individual instructions for all the cores in the cluster. To counteract this effect, in the Dustin cluster we introduce support for a novel Vector Lockstep Execution Mode (VLEM), where all cores execute the same instructions cycle-by-cycle.
When the cluster is configured in VLEM, core 0 acts as a \emph{leader} core, and the other 15 act as \emph{followers}. While VLEM is active, the IF stages and private caches of the \emph{follower} cores are clock-gated to save energy, and only the \textit{leader} core fetches the instruction and forwards them to the \emph{follower} cores. Fig.~\ref{fig:vlem_pic} provides a high-level overview of the two systems. 
To enter or exit VLEM, all cores have to \textit{i}) synchronize on a barrier, and \textit{ii}) write to a memory-mapped register.

It is crucial to make sure that all cores are in sync when entering VLEM and stay in sync during VLEM execution. Load/store operations simultaneously accessing the same TCDM bank are a potential source of desynchronization. In MIMD mode, the TCDM interconnect via round-robin gives access to the cores, delaying multiple accesses to the same bank. The core's memory accesses are carried out using a request/grant handshake: if the grant is not asserted, the core stalls until it arrives.
To keep cores aligned in VLEM, this mechanism is extended to hold all grant signals until all memory accesses have been completed.
The grants are then released simultaneously, preserving the synchronization.

\begin{figure}[t]
\centerline{\includegraphics[width=0.5\textwidth]{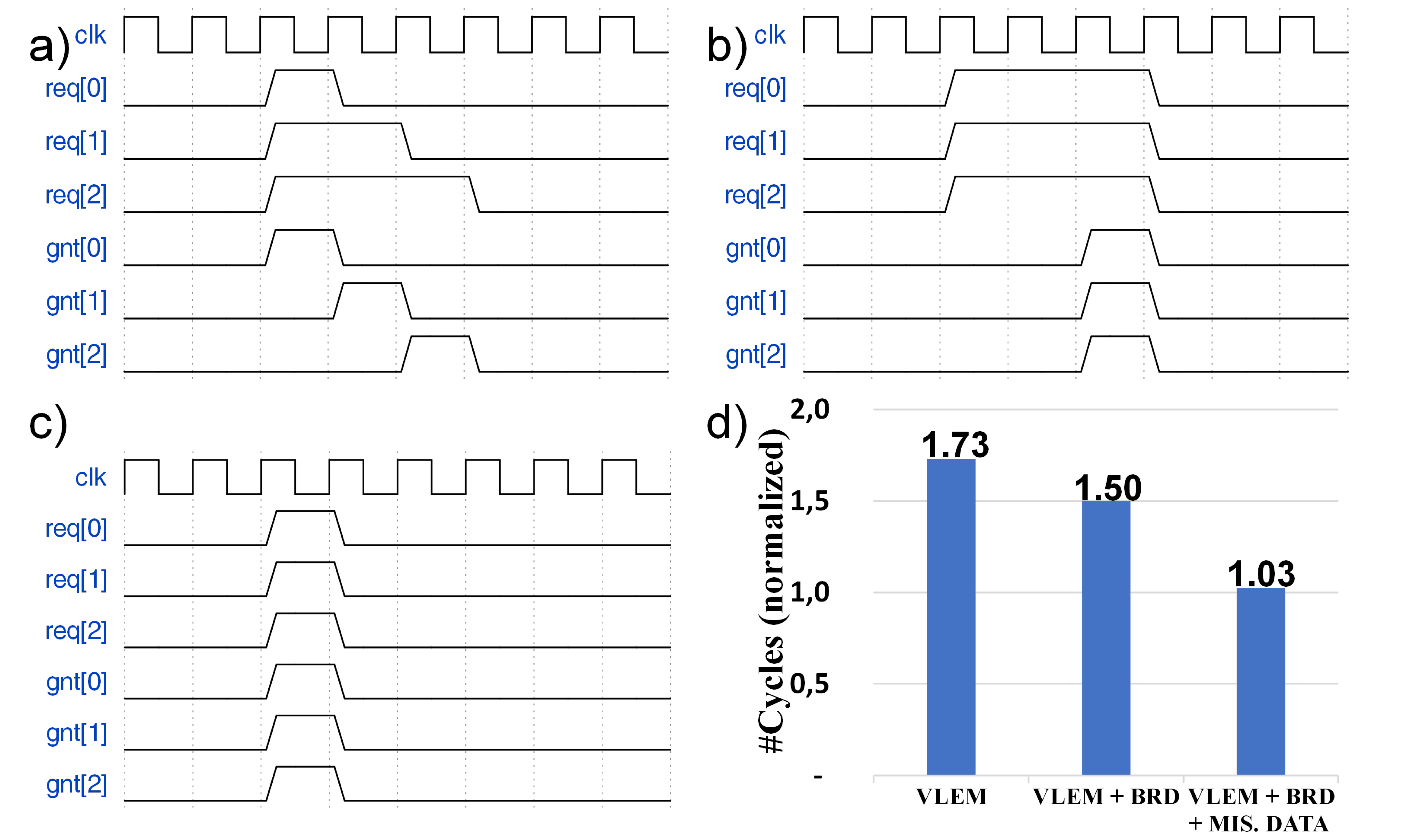}}
\caption{Request/Grant handshake for memory accesses at the core's interface. All cases present 3 concurrent memory requests on the same bank. a) MIMD case, sequentially served; b) VLEM case, sequentially served but Grant held by LKS unit for synchronization; c) Same bank and same address in VLEM which trigger broadcasting and resolution in one cycle; d) Performance overheads for different VLEM optimization implementation.}
\label{fig:someVLEMwaves}
\vspace{-0.5cm}
\end{figure}

Fig.~\ref{fig:someVLEMwaves} shows a simplified example of this behavior with 3 cores. In MIMD mode (Fig.~\ref{fig:someVLEMwaves}(a)), the 3 cores try to access the same bank simultaneously. The request signal of the three cores is asserted by all concurrently, but the grant is given sequentially, starting from core 0 to 2. Whenever the grant is received, the core can keep executing the rest of the code; otherwise, it stalls until the grant is received.
In VLEM (Fig.~\ref{fig:someVLEMwaves}b), the starting point is similar, but the grant for cores 0 and 1 are held until core 2 can also be served.
As all memory accesses are served simultaneously, the cores remain synced.

Keeping cores always in sync can lead to performance overhead whenever all cores repeatedly access the same bank. In MIMD mode, this is not an issue - the mechanism described in Fig.~\ref{fig:someVLEMwaves}a desynchronizes them, which means the overhead is typically only paid once. In VLEM, however, without specific countermeasures, the cores would hit serialization overhead in all successive accesses -- a penalty from 2 to 16 cycles per access depending on the number of conflicts. Fortunately, a common case is that of cores accessing the same word in the same bank, e.g., a pointer to the base of a shared array, such as a weight filter utilized by all cores in a DNN.
To avoid any overhead in this instance, we employ a hardware \textit{broadcasting} mechanism in Dustin.
It works by snooping the addresses of memory loads from all cores, comparing them, and propagating to memory single access if they are equal. The value extracted from memory is broadcasted to all the cores allowing 16 data accesses with one request.

Broadcasting does not solve conflicts occurring when accessing the same bank at different offsets, happening when memory structures with a size multiple of the number of banks are used. For instance, let us assume to have an output tensor of 8-bit data with a $H\times W\times C$ layout with dimensions $16\times 16\times 16$. Supposing that the tensor starts from an address aligned with the banks (Dustin includes 32 banks of 32-bit words, hence aligned to 128 bytes), core 0 will access bank 0 at address 0; core 1 will access with an offset of $16\times 16\times 16 = 4096$ bytes that also aligned to 128 bytes (bank 0); core 2 and the rest will all be accessing multiple of 4096 bytes, meaning that all cores will access the bank sequentially.
In highly arithmetic-intensive kernels, these conflicts are typically systematic: as the cores proceed in sync, they consistently access the same bank due to word-level bank interleaving.

A simple yet effective software-based countermeasure is to allocate data so that the base addresses seen by each core are never aligned in the same bank. Section~\ref{sec:software} discusses this problem and the related software countermeasures in detail.

In Fig.~\ref{fig:someVLEMwaves} d) we report the performance overhead of VLEM compared with the baseline case of execution in MIMD mode using a CNN layer as an example. Without any countermeasure, the number of cycles is increased by 73\%, translating into meager performance. 
Broadcasting reduces the overhead to 50\%, which is still not acceptable.
By adding ad-hoc data misalignment, so that base addresses are never on the same bank, overhead is down to 3\% in this specific test.

\begin{figure}[t]
\centerline{\includegraphics[width=0.5\textwidth]{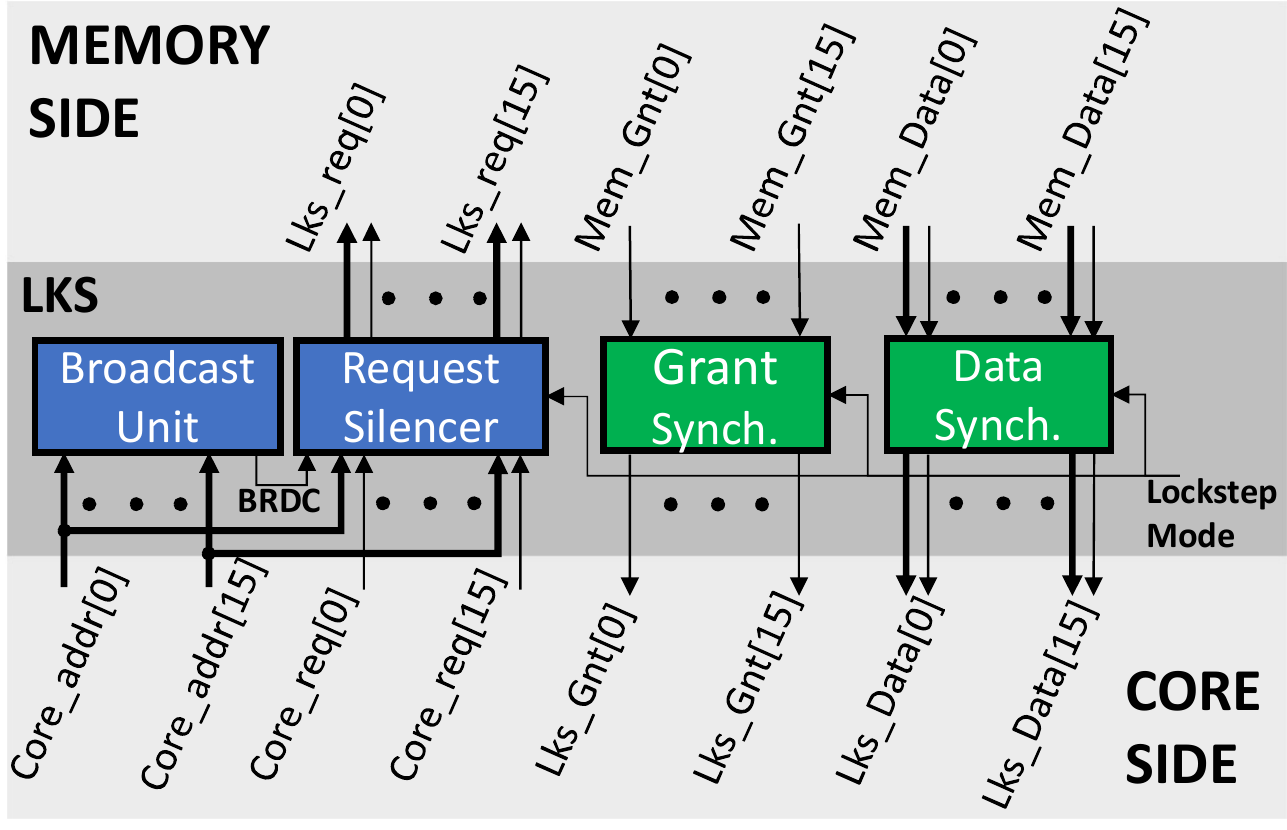}}
\caption{Internal components of the Lockstep Unit that sits between the cores and memory.}
\label{fig:lks_unit}
\vspace{-0.5cm}
\end{figure}

VLEM can be activated/deactivated in a single cycle by setting a memory-mapped register. When VLEM is engaged, the lockstep unit (LKS) is activated.
Fig.~\ref{fig:lks_unit} shows the key components of the LKS. Blue blocks indicate the path from the cores to memory, while green blocks show the return path. In MIMD mode, the LKS is bypassed.
In VLEM, the \emph{Broadcast Unit} filters the memory requests. The unit matches request addresses and asserts a broadcast signal (\textit{BRCD}) to the \emph{Request Silencer}. The \emph{Requests Silencer} then blocks requests from the \emph{follower} cores and only forwards the \emph{leader's} one.
The \emph{Grant Synchronizer} and \emph{Data Synchronizer} are responsible to keep the cores synchronized in terms of conflict on the memory banks, and when a broadcasting event happen, they are also responsible to forward Data to the \emph{follower} cores.

Entering VLEM also has an impact on the instruction cache subsystem. Specifically, the \textit{leader} core's private cache is not touched, but the caches of \emph{follower} cores terminate any in-flight cache refill and then enter sleep.
On exit of VLEM, the follower's PC will be set to the \textit{leader} value, and the caches will be turned on again.
Depending on the number of instructions executed during VLEM, switching back to MIMD mode can cause the \emph{follower} cores to stall immediately due to instruction cache misses.

\begin{figure}[t]
    \centering
    \includegraphics[width=\linewidth]{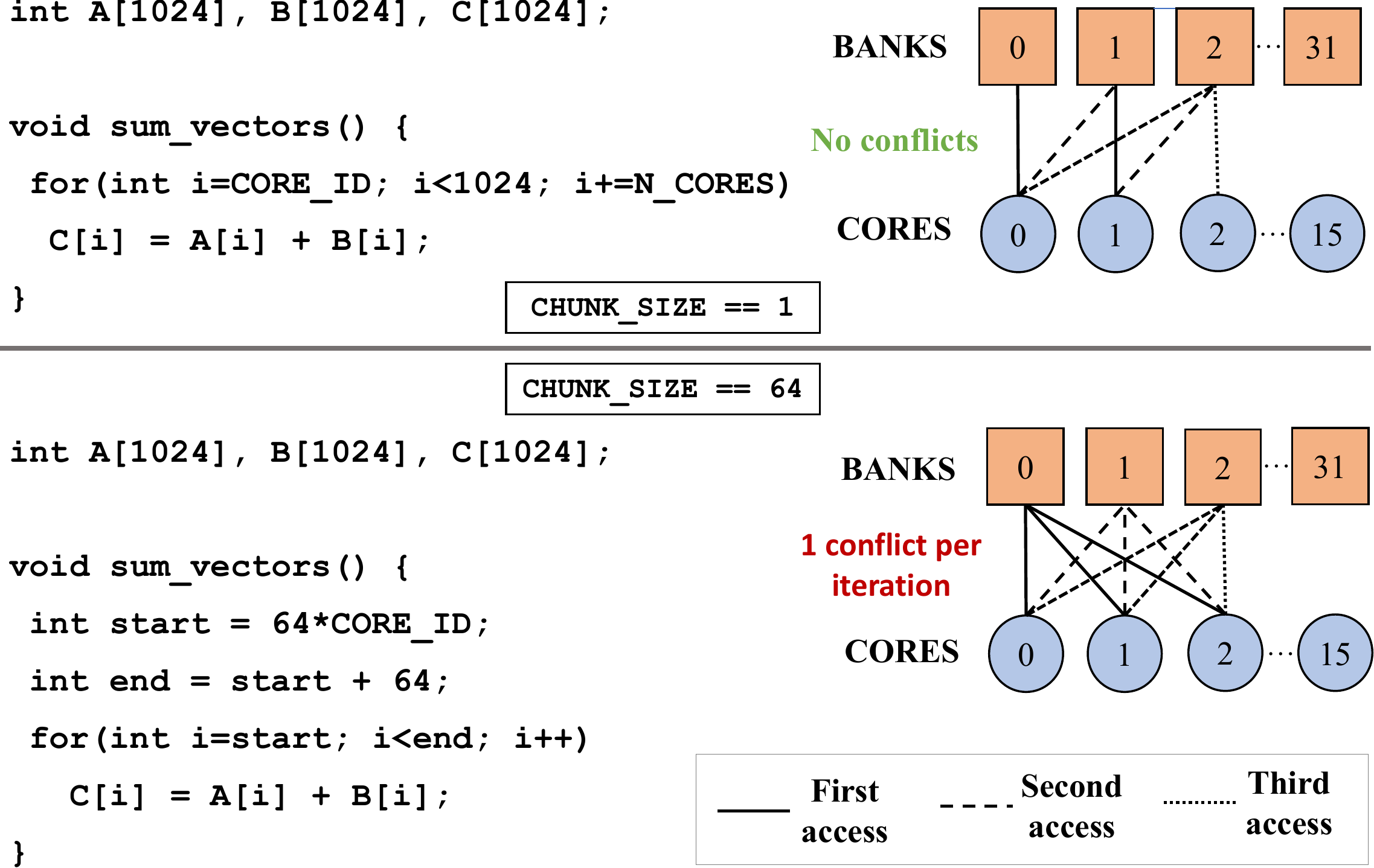}
    \caption{Element-wise vector addition with chunk size equal to 1 (top) and 64 (bottom). Access to the same memory bank by different cores executing the same instruction results in systematic bank conflicts.}
    \label{fig:software1}
\vspace{-0.5cm}
\end{figure}

\section{Software stack and programming model}
\label{sec:software}
We add dedicated extensions to the PULP GCC compiler and its hardware abstraction layer (HAL) to support the specific features of Dustin (i.e., multiple-precision arithmetic and VLEM). We define a set of \emph{intrinsics} in the GCC backend for each mixed-precision variant of MAC and dot-product operations. SIMD vectors do not comply with the GNU vector extension but are represented as 32-bit opaque types (i.e., \emph{int32\_t}). This design choice is motivated by the fact that GCC does not handle data formats of less than eight bits, while we need support for 2-bits and 4-bits elements. Nevertheless, this approach is totally transparent from the user perspective; as the only limitation, the programmer does not benefit from static type checking for multi-precision arithmetic.
In addition to the compiler backend, we extend the HAL with additional support to facilitate VLEM programming. The baseline HAL provides a set of primitives for core identification, synchronization, and memory allocation.

Core identification is achieved through a function that returns the unique identifier (\emph{core\_id}) of the core; on the Dustin platform, the $core\_id$ is an integer value in the range $[0 .. 15]$. Programmers can exploit loop-level parallelism, partitioning loop iterations into chunks and distributing these chunks to the executing cores in a round-robin order. This approach needs to include core identifiers into the loop control expressions (i.e., initialization, condition check, and increment). Core identification is extensively used in both modes (i.e., MIMD and VLEM).
The primary synchronization primitive provided by the HAL is called \emph{barrier}. This function stops a core until all other cores execute an associated barrier function, enforcing a synchronization point in the program flow. Programmers must explicitly add a call to this function to guarantee data consistency between adjacent code regions. However, the event unit component \cite{glaser2020energy} provides low-overhead synchronization support, enabling power-saving policies for waiting cores. The barrier is the main synchronization mechanism adopted in Dustin during MIMD operation.
\begin{figure}[t]
    \centering
    \includegraphics[width=\linewidth]{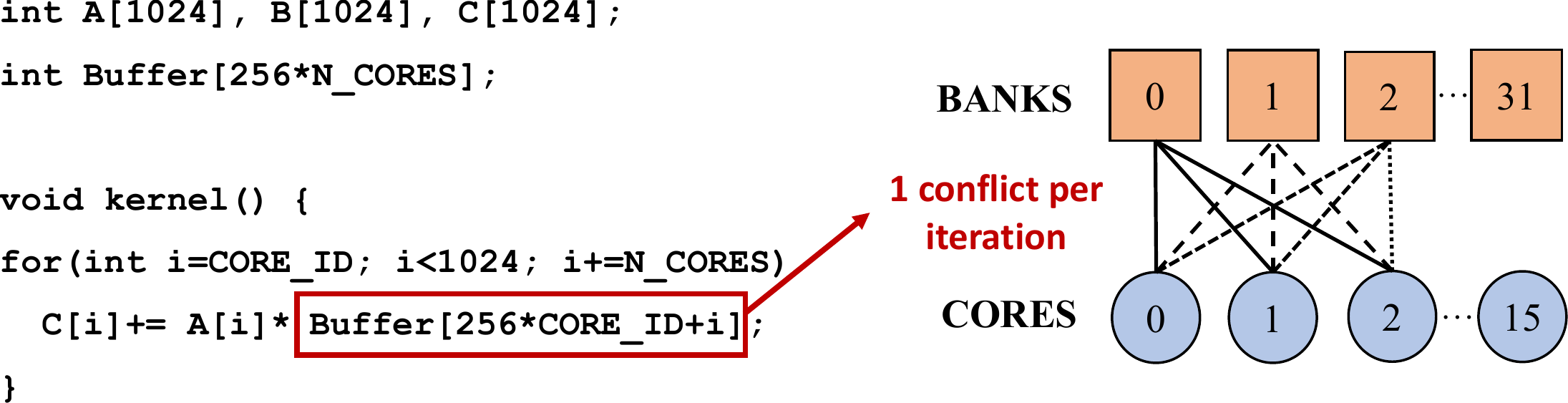}
    \caption{Code template for an application kernel accessing an intermediate buffer with an expression depending on $core_id$. This pattern results in systematic bank conflicts when the buffer size is a multiple of the number of banks multiplied by the word size.}
    \label{fig:software2}
\vspace{-0.5cm}
\end{figure}
\begin{figure}[t]
    \centering
    \includegraphics[width=0.6\linewidth]{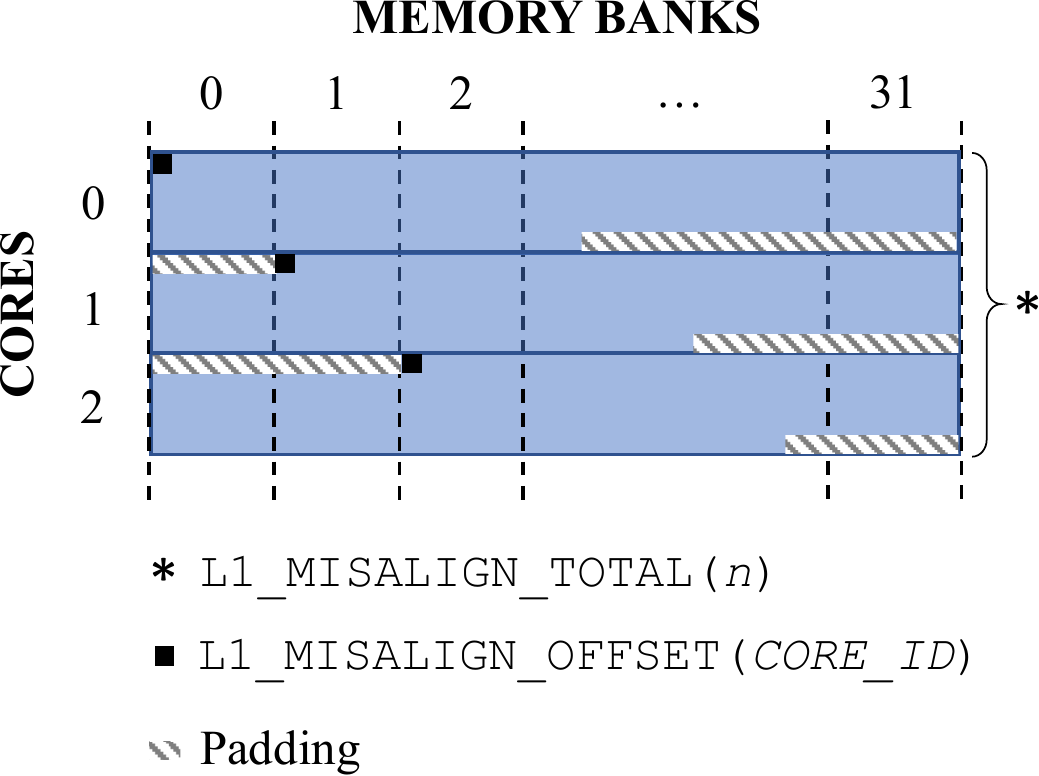}
    \caption{Memory layout of a statically allocated global buffer, highlighting the offsets of the core sub-buffers and the padding areas required to avoid bank conflicts.}
    \label{fig:software3}
\vspace{-0.5cm}    
\end{figure}
Finally, the original PULP HAL supports static and dynamic memory allocation. Static allocation requires specifying the data size at compile time through constants (e.g., the sizes specified for arrays must necessarily be constant values). In C++ programs, \emph{constexpr} expressions can replace constant literals; anyhow, their result is constant and computed at compile time. Global variables can be placed in TCDM or L2 memory areas decorating their declaration with a preprocessor macro (PI\_L1 or PI\_L2, respectively) mapped on a \emph{\_\_attribute\_\_((section(...)))} directive. Following C and C++ semantics, local variables are on the stack; as discussed later, the stack memory area is allocated on the L1 memory.
Dynamic allocation on the heap memory area is available through HAL primitives using a standard heap allocator based on malloc/free functions. The malloc function requires specifying the size of the memory area to allocate and returns a pointer of type \emph{void} that can be cast into a pointer of any form. The \emph{free} function de-allocates a memory region, which becomes available for the following malloc calls. The HAL provides alternative malloc/free functions to allocate data in L1 or L2 memory.

From the memory access perspective, algorithms can adopt two alternative memory access patterns, \emph{strided} or \emph{indirect}. The strided pattern consists of a regular sequence of accesses characterized by an initial address, a distance between adjacent accesses (called \emph{stride}), and the number of accesses in the sequence. E.g., a sequence can be expressed as $A[i]_{0:1:N-1}$, where A is an array variable, and i is a loop induction variable ranging from $0$ to $N-1$ with stride $1$. The indirect pattern is a sequence of accesses where multiple memory requests are required to access each element.

\begin{figure}
    \centering
    \includegraphics[width=0.95\linewidth]{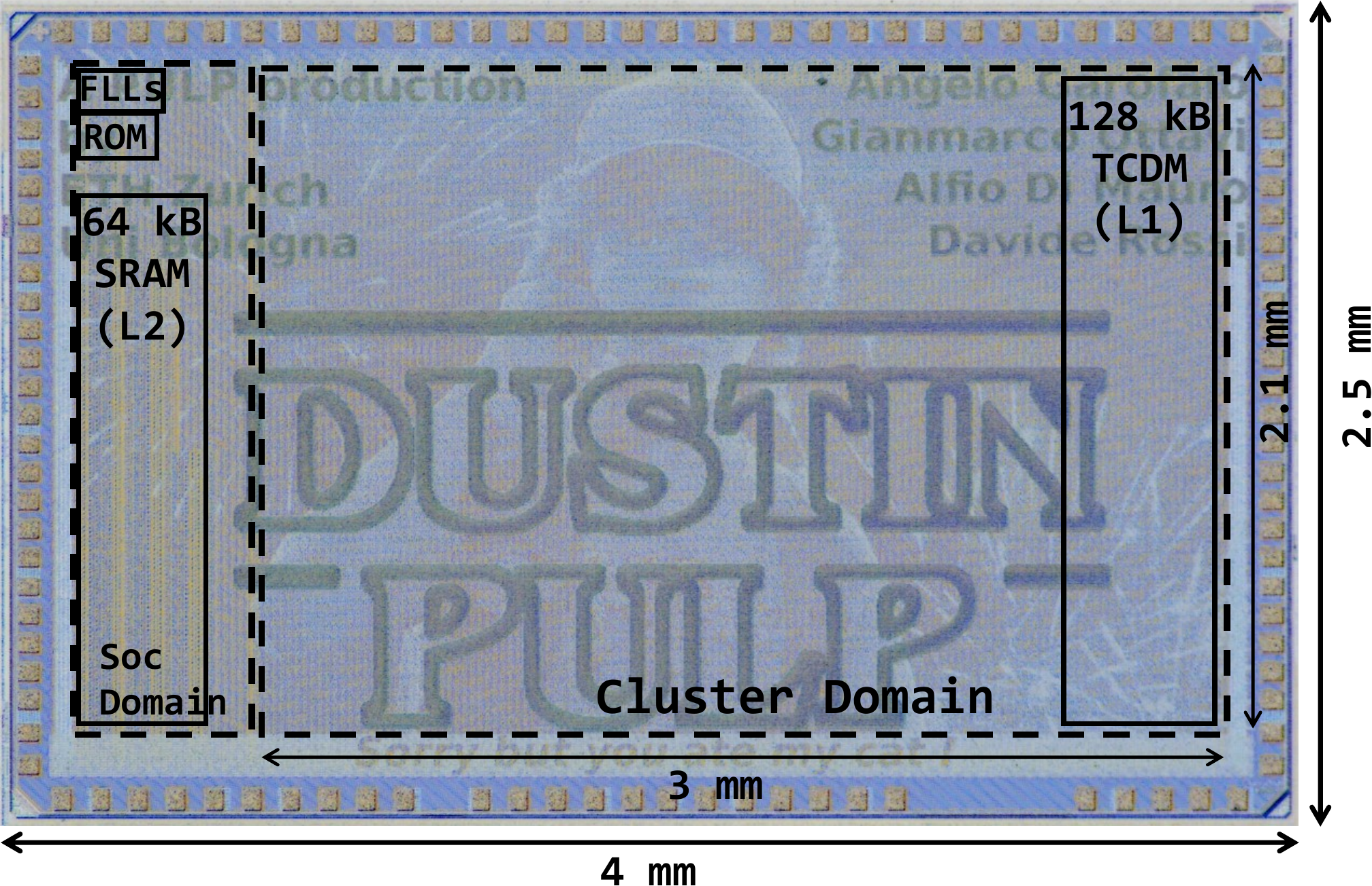}
    \caption{Chip micrograph.}
    \label{fig:floorplan}
\vspace{-0.5cm}
\end{figure}
\begin{table}[t]
\caption{Detailed area breakdown of Dustin compute cluster}
\label{tab:dustin_ref_area}
     \centering
     \resizebox{\linewidth}{!}{
     \input{area_breakdown_table}     }
\end{table}
\begin{figure}[t]
    \centering
    \includegraphics[width=\linewidth]{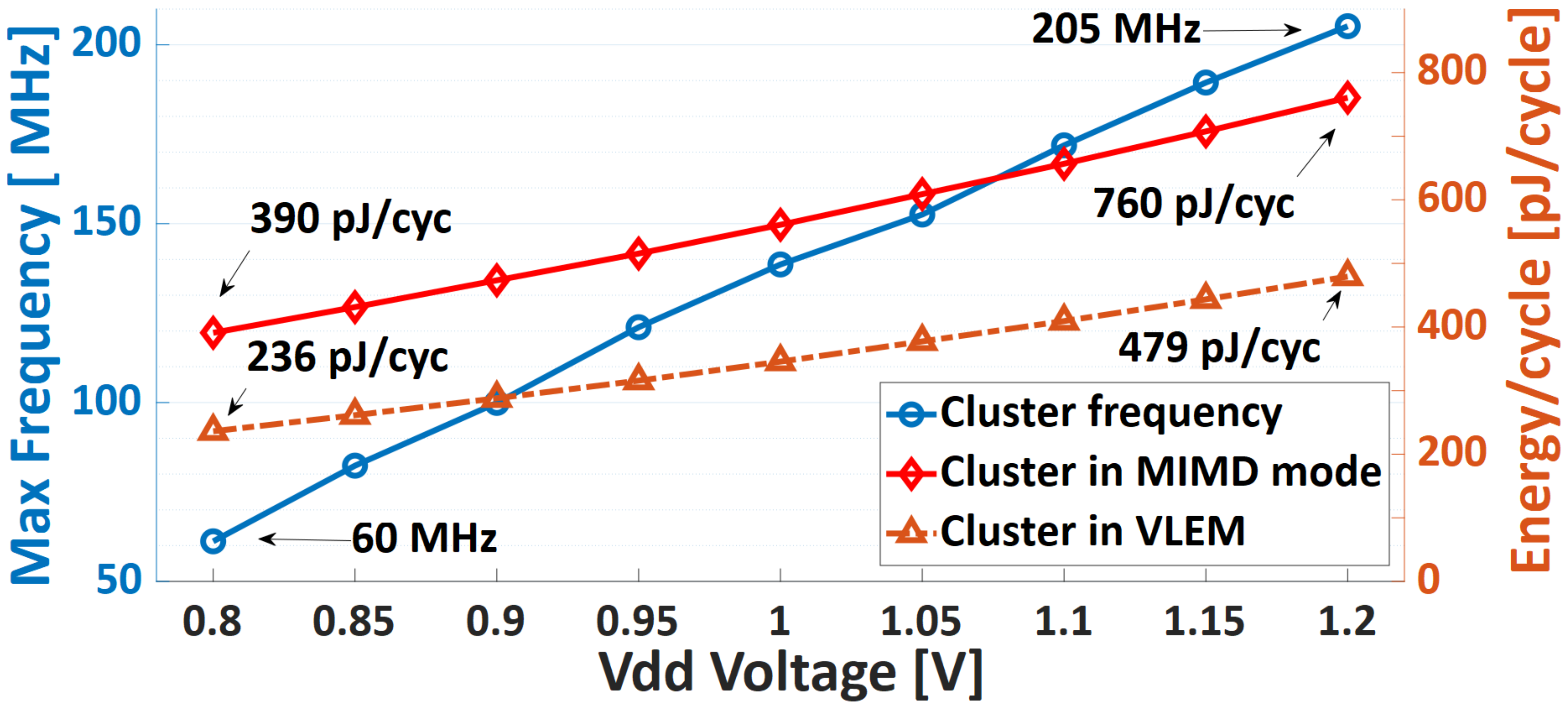}
    \caption{Voltage Sweep vs. Max Freq. vs. Energy/Cycle.}
    \label{fig:sweep-vdd-fmax}
\vspace{-0.5cm}
\end{figure}
\begin{figure}[t]
    \centering
    \includegraphics[width=0.9\linewidth]{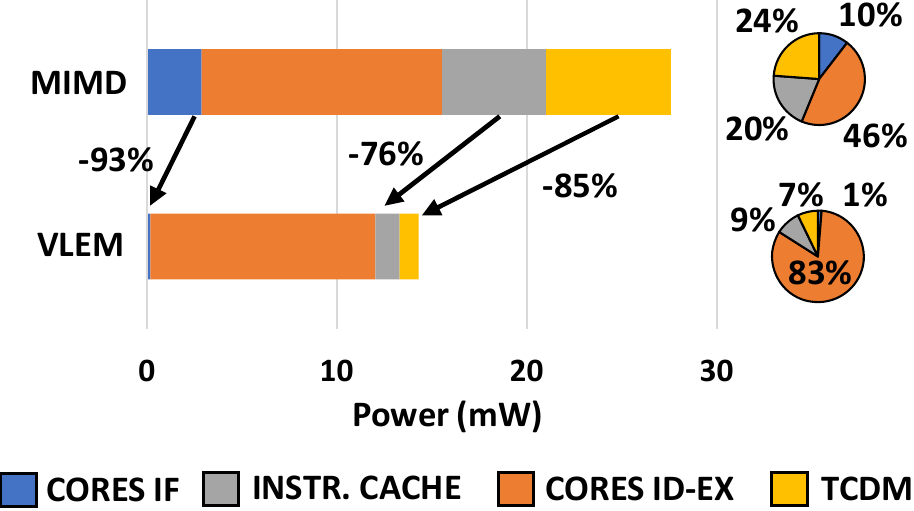}
    \caption{Power in an inner loop of a CNN layer (MIMD vs. VLEM).}
    \label{fig:power_breakdown}
\vspace{-0.5cm}
\end{figure}

An analysis of the memory access pattern becomes critical when VLEM is active because some cases can induce systematic bank conflicts that are detrimental to performance, as shown in the upper part of Fig.~\ref{fig:software1}. In the case of a stridden access pattern, programmers can avoid bank conflicts by coalescing memory accesses at the bank level. This property holds if adjacent cores execute the same instruction access addresses located on contiguous banks. The actual feasibility of this technique does not depend exclusively on the memory access pattern but also on the loop parallelization strategy. Fig.~\ref{fig:software1} illustrates a code performing element-wise vector addition for two alternative chunk size values at the two extremes of the possible range (i.e., 1 and 64). The access pattern has a unit stride, but the absence of bank conflicts depends on the chunk size. Setting the chunk size to 1 prevents bank conflicts, while the value 64 induces all the accesses to insist on the same bank.
In the general case, if $s$ is the byte size of the array element (less or equal to 32 bits), $w$ is the word size, $n_C$ is the number of cores, and $n_B$ is the number of banks, a chunk size in the range between $w/s$ and $(n_B/n_C) * (w/s)$ avoids the presence of bank conflicts. The Dustin HAL provides the macros \texttt{MIN\_CHUNK\_SIZE} and \texttt{MAX\_CHUNCK\_SIZE} to retrieve these values specifying only the element size since the other parameters are architecture-dependent.

Fig.~\ref{fig:software2} shows a code template presenting an access pattern that is dependent on the core ID. This pattern is commonly found in libraries that perform operations on intermediate buffers, such as PULP-NN~\cite{garofalo2020pulp}. The occurrence of this pattern can result in bank conflicts, depending on the starting address of the individual buffer regions. These conflicts are systematic when the buffer size is a multiple of $w * n_{B}$. In the general case, conflicts occur if the set of remainders between the starting addresses referred to by each core and $w * n_{B}$ does not contain exactly $N_{C}$ elements. To address this issue, the Dustin HAL provides a function, called \texttt{l1\_misaligned\_malloc}, for dynamic memory allocation of intermediate buffers on L1.

\begin{figure}[t]
\centerline{\includegraphics[width=\linewidth]{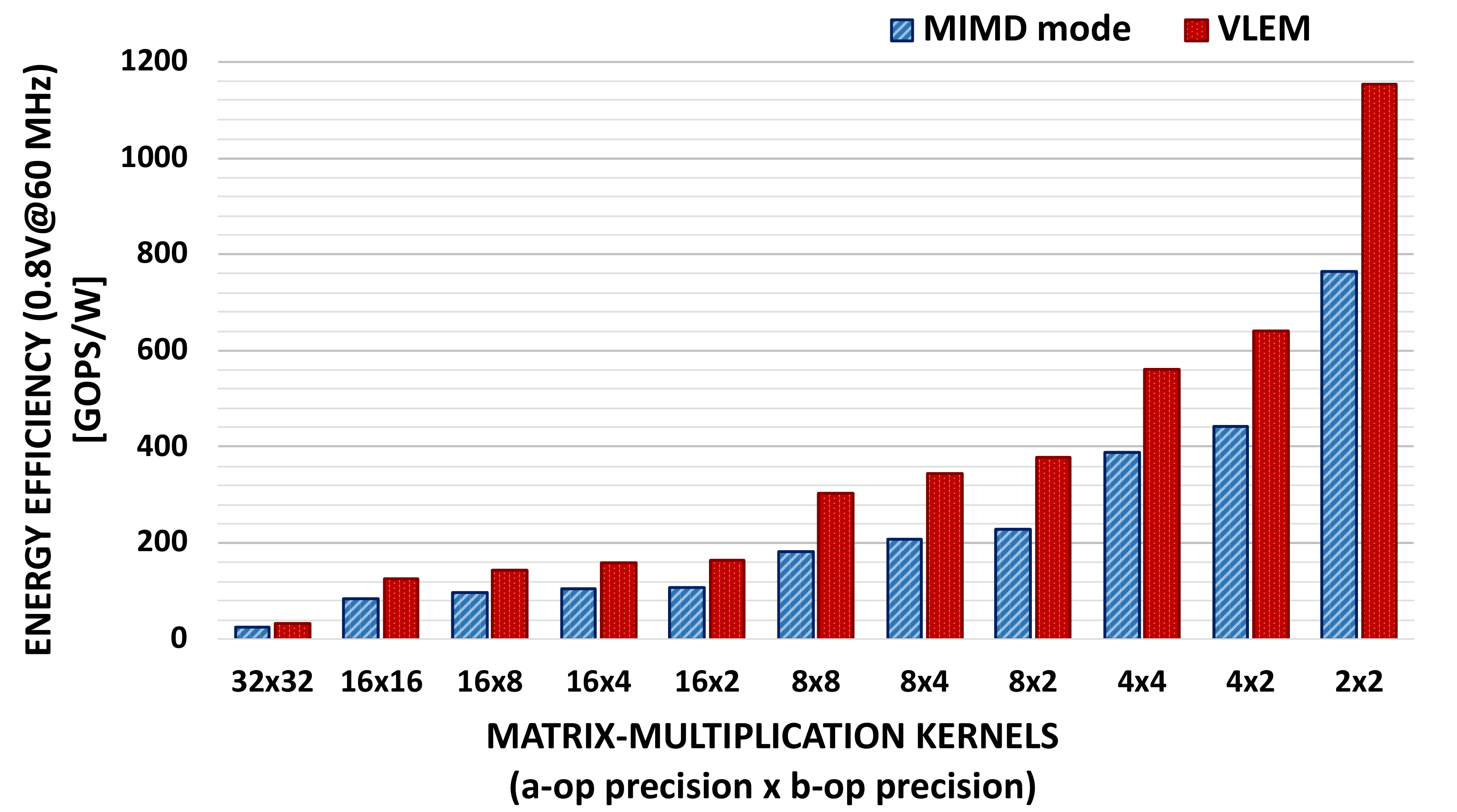}}
\caption{Comparison in terms of Energy Efficiency of Dustin configured in MIMD and VLEM, running Mixed-precision Matrix Multiplication kernels.}
\label{fig:ee-pic}
\vspace{-0.5cm}
\end{figure}

This function is a wrapper around the malloc operations, forcing the starting address of the allocated memory space to be allocated on the bank number equivalent to the $core\_id$. This approach enables each core to allocate its local buffer with a strong guarantee of the absence of bank conflicts in the case of strided access patterns.

Some application domains strictly require static allocation of a single memory buffer shared among the cores, precluding the use of \texttt{l1\_misaligned\_malloc}. In these cases, the Dustin HAL provides a macro \texttt{L1\_MISALIGN\_TOTAL} that computes the total amount of contiguous memory required to fulfill requirements and avoid structural bank conflicts. Additional padding space is introduced to move the start address of each core on a different bank.
Supposing that each core requires $n$ bytes, the macro returns $n_C * sub\_buffer\_size$, with $sub\_buffer\_size = \lceil (n + (n_C * w))/(n_B * w) \rceil * (n_B * w)$. In addition, the macro \texttt{L1\_MISALIGN\_OFFSET} returns the starting offset of the sub-buffer related to $core\_id$ as $core\_id * sub\_buffer\_size + core\_id * w$.
Fig.~\ref{fig:software3} illustrates the memory layout of a statically allocated global buffer. The padding area between adjacent sub-buffers varies based on the value $n$, and it is equal to $n_B * w - 1$ bytes in the worst case. 
Once misalignment is achieved, the performance benefits are shown in Fig.~\ref{fig:someVLEMwaves}(d), in particular, going from the middle to the rightmost bar. This outcome is unrelated with the shape of the array being computed, which only impacts the memory requirements. For instance, a 3D Convolution with 16 input channels and a 3x3 filter necessitates an input buffer of 6144 bytes as opposed to 4608 bytes (as calculated using the above formula).

\section{Chip Design and Silicon Measurements}

Fig.~\ref{fig:floorplan} shows a die photograph of Dustin. The SoC is implemented in TSMC 65nm CMOS technology targeting a clock frequency of 200 MHz in typical operating conditions, within a die size of 10 $mm^2$. In the following, we analyze the measurements of Dustin's cluster, leaving aside the measurements of the SoC subsystem used in this work as a software programmable testbench hosting the RISC-V core, a standard set of peripherals, and 80 kB of SRAM memory.

%
Fig.~\ref{fig:sweep-vdd-fmax} reports Dustin's cluster maximum operating frequency and energy per cycle at different supply voltages, ranging from 0.8~V to 1.2~V. The measurements are carried out on the silicon prototype, running matrix multiplication kernels with 8-bit precision operands, a typical high-utilization deep neural network workload. The operating frequency increases with the voltage up to the maximum of 205 MHz, measures with an operating voltage of 1.2V. In terms of energy, we notice a significant saving factor when the cluster runs a matrix multiplication kernel in VLEM, about 38\% lower energy per cycle compared with the MIMD mode in all the voltage corners considered. This result on regular kernels like matrix multiplications is achieved thanks to the clock gating applied to the caches and IF stages of the \emph{follower} cores, which are not used in VLEM, reducing the cluster dynamic power consumption.

\begin{figure}[t]
\centerline{\includegraphics[width=\linewidth]{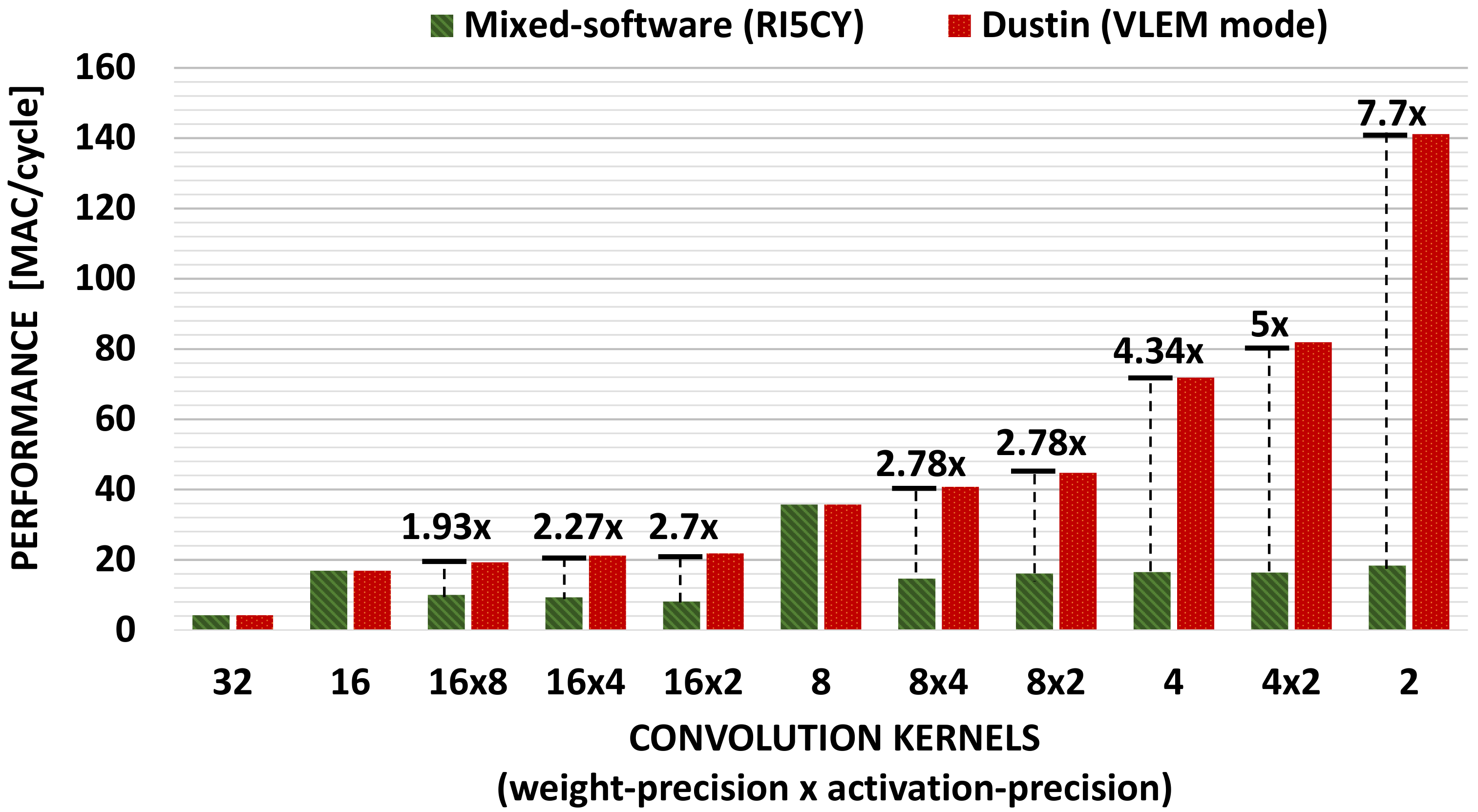}}
\caption{The chart compares the execution of mixed-precision convolution kernels running on the baseline 16 cores cluster with the RI5CY core (software mixed-precision kernels) and on Dustin's cluster in VLEM (featuring the Mixed-precision ISA extensions).}
\label{fig:gops-pic}
\vspace{-0.5cm}
\end{figure}

To understand the costs in terms of area and power of our contributions, we implement a baseline version of the cluster, stripped of VLEM logic and the Mixed-precision capabilities, with a full synthesis and place and route flow. We compare the two clusters from a physical point of view and from a performance and energy efficiency perspective. Table~\ref{tab:dustin_ref_area} reports the comparison between the two clusters in terms of area, detailing the breakdown of the main components of the cluster and the core.
The Dustin cluster is 5\% bigger than the reference one. The main contributors to this overhead are the cores, where most of the new logic resides. The \emph{leader} and the \textit{follower} cores show the same area, since the \textit{follower} cores still feature IF stages to be used in MIMD mode. The additional wires of the \emph{leader} core to broadcast the instructions to the other cores (when the cluster is in VLEM) contribute with negligible overhead. 

The features added to Dustin’s cores brought a 17\% of area overhead when compared to RI5CY. The IF stage has been modified to incorporate the new logic to enable the VLEM, while ID includes a mixed-precision controller and modifications to the decoder to support the new set of virtual instructions. The EX stage contributes the most to the core area increase at 28\% (w.r.t. RI5CY ex-stage), as it features a new extended DOTP unit designed to handle 4-bit, 2-bit, and all mixed precision permutations between 16-bit and 2-bit. Additionally, the LSU area has increased, as the TCDM interconnect requires increased driving strength for the increased complexity brought by the LKS. This effect is due to the lockstep and broadcast units along the path from the cores to the TCDM banks. Finally, the CSR includes new registers to store the format of SIMD mixed-precision instructions. Looking at the whole Dustin’s cluster, each core has a modest contribution in area of only 2.1\%.

In conclusion, the proposed approach demonstrates an affordable 5\% increase in area overhead with an impact of 13\% timing wise, which takes into account the broadcast mechanism. We argue that these trade-offs are acceptable given the significant improvements in terms of performance and energy efficiency that the mixed-precision and VLEM features bring over the baseline cluster.

To fortify the previous statement, Fig.~\ref{fig:power_breakdown} reports the power breakdown of the innermost loop of a QNN convolution layer. The values shown are the results of post-layout simulations running Dustin's cluster at 50 MHz, at the supply voltage of 1.0V, in the typical corner (TT, 25C). We run post-layout simulations because such a study would not be possible from silicon measurements since all the cluster components share the same power lines, and no fine-grained breakdown can be extracted.
In the convolution kernel considered, which follows the data flow presented in~\cite{garofalo2020pulp}, each core of the cluster process different subsets of the input feature map over the same set of weights to produce different sub-sets of the output feature map. When the cluster runs in VLEM, such a layout allows to massively leverage Dustin's broadcast features on the QNN weights, significantly reducing the power consumption of the TCDM interconnect, in addition to the clock gated caches and IF stages of the \emph{followers} cores. In the pie chart on the right of Fig.~\ref{fig:power_breakdown}, we stress the fact that most of the power is spent on computation (ID-EX) in VLEM, eliminating the overhead of moving back and forth the same data from the TCDM to the cores and independently fetching the same instructions for all the cores, as it happens instead in MIMD mode.

\subsection{Benchmarking}
To highlight the performance and the energy efficiency of the silicon prototype on QNN workloads, we firstly benchmark heavily quantized and mixed-precision convolution kernels, varying the format of the operands from 2-bit to 16-bit, covering all the relevant operand precision permutation scenarios. Then, we present the performance of the proposed SoC on an end-to-end DNN for TinyML applications.

\begin{figure}[t]
\centerline{\includegraphics[width=\linewidth]{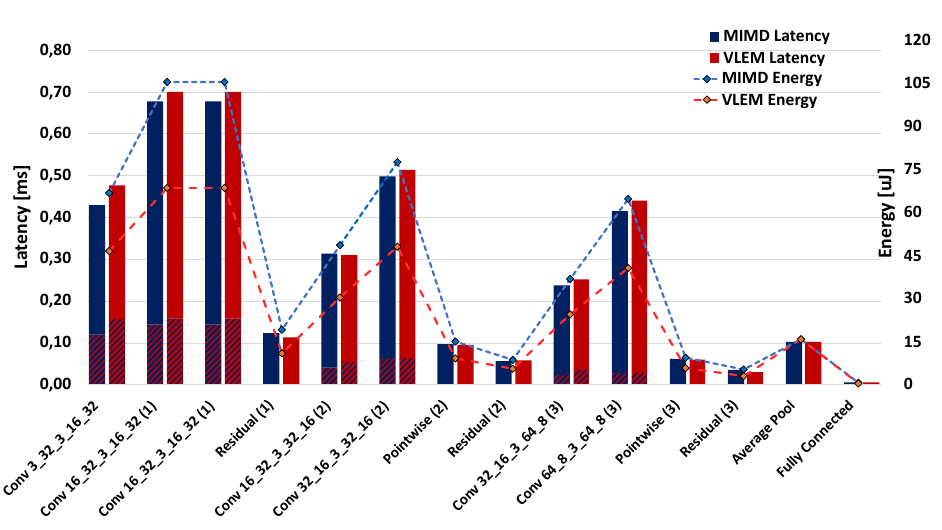}}
\caption{Latency and Energy Consumption of ResNet8 executed in MIMD and VLEM modes on the Dustin cluster. Textured bars indicate time spent in the IM2Col phase of the layer (executed in MIMD). The rest of the time is time spent on MatMul (executed in VLEM).}
\label{fig:resnet8_mimd_vlem}
\vspace{-0.5cm}
\end{figure}

\subsubsection{Quantized DNN Kernels}

In Fig.~\ref{fig:gops-pic}, we report the performance of the kernels running on Dustin's cluster, and we compare the results with the baseline cluster described above, featuring the RI5CY cores. In the latter case, sub-byte and mixed-precision kernels are handled purely in software, as shown in~\cite{bruschi2020enabling}. We notice that on kernels where only the activations are sub-byte operands, the performance benefits of the hardware support for mixed-precision computation range from 1.9$\times$ to 2.8$\times$ due to the unpacking functions used in RI5CY, but in a less arithmetic intensive portion of the kernel. In all other configurations, the mixed-precision ISA extensions provide a significant advantage ranging from 2$\times$ to 7.7$\times$ improvements w.r.t. the baseline cluster, where the unpacking operations must be performed even in the innermost loop of the convolution, degrading the performance heavily. 

To highlight the energy savings of the VLEM on regular computing kernels, we measure the energy consumption with the cluster running the matrix multiplication in two modes: the classic MIMD mode and the VLEM (enabled via software). Fig. \ref{fig:ee-pic} shows the related efficiency. The execution of linear kernels in VLEM achieves 1.5$\times$ better energy efficiency and no performance overhead w.r.t. the default MIMD execution. 

\begin{figure}[t]
\centerline{\includegraphics[width=\linewidth]{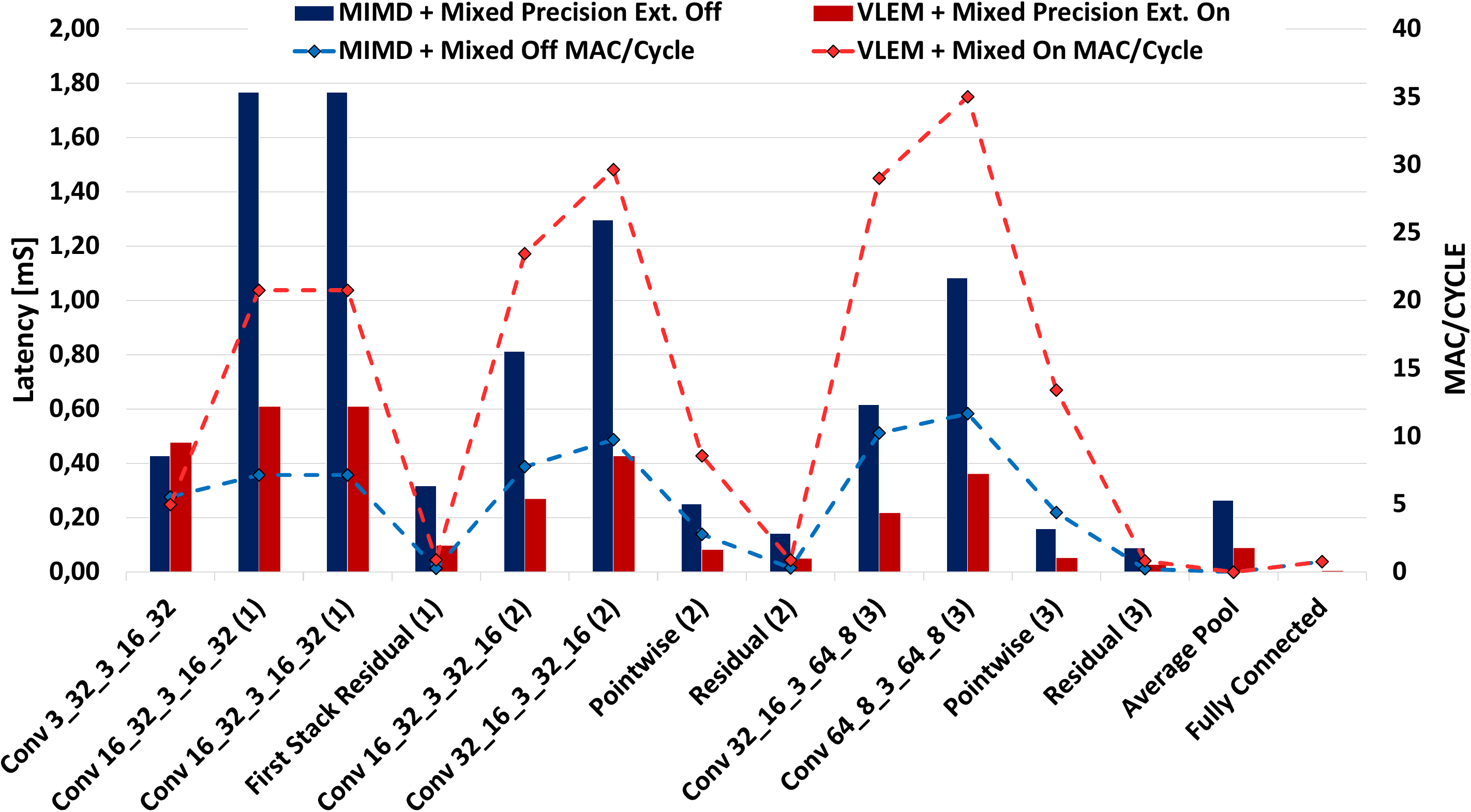}}
\caption{Execution time and computational efficiency of ResNet8 quantized with 8-bit Activation and 4-bit Weights executed on a Dustin Cluster in VLEM mode with Mixed-Precision extensions off (blue bars) and on (red bars). }
\label{fig:resnet8_mixed}
\vspace{-0.5cm}
\end{figure}

\begin{table}[t]
\caption{ResNet8, comparison With State-Of-The-Art}
\label{tab:resnet-table}
     \centering
     \resizebox{0.45\textwidth}{!}{
     \input{resnet_table}
     }
    \begin{tablenotes}
      \item Comparison with other SoA Microcontrollers on the Restnet8 bechmarks.
    \end{tablenotes}
\end{table}

\subsubsection{ResNet8 Inference}
This section presents the results of the inference of a ResNet8 from the MLPerf Tiny Benchmark~\cite{banbury2021mlperf}, the leading benchmark suite for tiny devices at the edge. The results published online\footnote{https://mlcommons.org/en/inference-tiny-10} are trained on the CIFAR-10 dataset where we achieved a top-1 accuracy of 88\% on float, 8-, 8x4-, 4-bit quantization. However, a reduction down to 84\% was observed on 2-bit quantization. For this simple dataset, the number of bits chosen for the quantization of activations and weights has a relatively minor effect on accuracy. However, this may not be the case for harder datasets and more complex models. As shown by Capotondi et al.~\cite{capotondi2020cmix}, using mixed-precision quantization can provide significant benefits, particularly on the MobileNetV1 model. The mixed-precision quantized network achieved the highest accuracy, as the full precision (8-bit) could not fit in the memory of the MCU, forcing the implementation to use a shallower network leading to a great loss of accuracy. In this work, we choose to show the inference on an 8-bit and 8x4-bit mixed-precision configuration to highlight the benefits of mixed-precision hardware support for end-to-end DNN inference.

Following the computational model adopted in the PULP-NN library \cite{garofalo2020pulp}, each convolutional layer first creates an \textit{IM2COL} buffer into L1 memory, reordering the 3D input feature map into a 1D vector in Height-Width-Channel (HWC) tensor layout. Then the \textit{MatMult} can be performed. VLEM can only be applied effectively to the \textit{MatMul} kernel since \textit{IM2COL} buffer creation mainly consists of byte reordering operations and bitwise operations (for sub-byte convolutions), featuring very irregular patterns and challenging vectorization.

\begin{table*}[t]
\caption{Comparison With State-Of-The-Art}
\label{tab:soa-table}
     \centering
     \resizebox{\textwidth}{!}{
     \input{Soa_comparison_xpulpnn}
     }
    \begin{tablenotes}
      \item \footnotemark[1]{} OPs = 1 8-bit (or 4-bit or 2-bit) MAC on MatMul benchmark unless differently specified.
      \footnotemark[2]{} Execution on SW programmable Core. \\ \footnotemark[3]{} Square brackets indicates power consumption in VLEM.
    \end{tablenotes}
\end{table*}

In general, to have an efficient execution (i.e., a high number of MACs/Cycle), the cluster should spend most of its execution time on matrix multiplication rather than on IM2COL. With an HWC tensor layout, the number of input channels is proportional to the number of iterations of the kernel inner loop. In contrast, the number of output channels is proportional to the times the inner loop is repeated. While the configuration characterized by 32 input channels and 64 output channels with a 3$\times$3 kernel used for synthetic explorations in the previous section reaches a peak performance close to the ideal (35 MAC/Cycle), this specific configuration is not present in the ResNet8. More specifically, the network input layer presents 3 input channels and 16 output channels featuring computational efficiency of 16.3\% (5 MAC/Cycle), where computational efficiency is defined as the ratio between ideal performance and actual performance. On the other hand, deeper layers present higher computational efficiency, closer to the ideal: 60\%; 75\% and 90\% for the first, second, and third group, respectively, as shown by the dotted lines in Fig.~\ref{fig:resnet8_mixed}.

Fig.~\ref{fig:resnet8_mimd_vlem} shows a comparison of execution time and the energy consumption of the network's layers between MIMD and VLEM modes at the operating frequency of 200MHz (1.2V). VLEM is applied to the MatMul kernel of all the layers except the last two.
In section \ref{sec:VLEM}, we mentioned the overhead that comes with entering and exiting VLEM. This is visible in the textured section of the graph (IM2COL). We can note that the more frequently we switch to VLEM, the larger will be the overhead (e.g., \textit{Conv\_3\_32\_3\_16\_32} requires 512 switches with an 11\% overhead, meanwhile, \textit{Conv\_32\_16\_3\_64\_8 (3)} only has 8 with 6\% overhead). Furthermore, a larger number of inner-loop operations helps hide the impact of entering and exiting from VLEM. Such a number is proportional to the number of input and output channels.
We can observe that, while the difference in execution time between the two modes of operation is negligible, the impact on energy is more significant. In general, the smaller the time spent in creating the \textit{IM2COL} over the overall layer time, the better the energy gain. This effect is also reflected in Fig.~\ref{fig:resnet8_mimd_vlem} where the dotted lines increase their distance when moving from left to right. Overall, the MIMD executes a frame in 3.72$ms$ spending 581$\mu J$; in VLEM, a frame is ready after 3.80$ms$ spending 373$\mu J$ saving ~35\% of energy at the expense of 2\% in execution time.

Finally, Fig.~\ref{fig:resnet8_mixed} shows the combined effect of VLEM and Mixed Precision extensions over a baseline cluster where both features are disabled. We use the ResNet8 network described before quantized with 8-bit activation and 4-bit weights, with the input layer and output layer quantized to 8-bit. This mixed-precision setting has been shown to achieve accuracy comparable to full-precision (8-bit) even on complex datasets~\cite{van2020bayesian}. We can note that exploiting a combination of VLEM and mixed-precision extensions on the full network reduces execution time by 63\% and energy consumption by 73\%. 
In Table~\ref{tab:resnet-table}, we present the results of peak performance for the specific benchmark that is available online. The Dustin processors were found to be ranked second and third in terms of latency and energy efficiency, with only processors utilizing dedicated accelerators surpassing their performance.

\subsection{Comparison with the state-of-the-art}

Table~\ref{tab:soa-table} shows a comparison with recently published IoT end-nodes and fully programmable clusters. Compared to traditional single-core IoT end nodes~\cite{SLEEPRUNNER,SAMURAI}, the proposed work delivers significantly better performance and efficiency thanks to the exploitation of parallelism. Compared to similar fully programmable multi-core IoT end-nodes~\cite{WOLF,VEGA}, implemented in 40nm and 22nm technology nodes, respectively, the proposed SoC delivers similar performance and energy efficiency on an 8-bit format, despite the less scaled technology node used for its implementation. The performance goal is achieved thanks to the larger parallelism of the cluster, which compensates for the less scaled node. Most significantly, if we compare the energy efficiency, we can note that it is also similar thanks to the VLEM execution mode saving up to 38\% of the overall power consumption of the cluster. 

If we compare Dustin and XPULPNN~\cite{garofalo2020xpulpnn} performance in uniform-precision kernels, we note a gap ranging from 30 to 20\%. This effect is related to \textit{MAC\&LOAD} instructions: they allow XPULPNN to compute MACs while loading the following operand simultaneously. Nevertheless, executing mixed-precision kernels adds the unpacking/packing overhead reducing the efficacy of \textit{MAC\&LOAD} instructions. In this scenario, Dustin firmly outperforms XPULPNN and reaches comparable energy-efficient results despite the substantial gap between the technologies used for implementation. It should be noted that Mixed-Precision extensions proposed in this work could be combined to \textit{MAC\&LOAD} to boost performance in both uniform mixed-precision quantized kernels.

\section{Conclusion}

We presented Dustin, a fully programmable Multiple Instructions Multiple Data (MIMD) cluster integrating 16 RISC-V cores featuring 2b-to-32b bit-precision instruction set architecture (ISA) extensions enabling fine-grain tunable mixed-precision computation. The cluster can be dynamically configured into a Vector Lockstep Execution Mode (VLEM), turning off all IF stages and L1 I\$ except one, reducing power consumption on average by 38\% with no performance degradation. The cluster, implemented in 65nm CMOS technology, achieves a peak performance of 58 GOPS and a peak efficiency of 1.15 TOPS/W -- competitive with IoT end-nodes using much more scaled and expensive technology nodes.

\bibliographystyle{IEEEtran}
\bibliography{main.bib}

\input{bio/bio.tex}

\end{document}

%% file: area_breakdown_table.tex
\begin{tabular}{lcccc}
\multicolumn{5}{l}{CLUSTER   LEVEL}                                                \\
                     & Dustin {[}µm$^2${]} & Ref {[}µm$^2${]} & Perc. & Delta{[}\%{]} \\ \hline
Cluster              & \num{3811874}         & \num{3607831}      & 100        & 5.66          \\ \hline
Icache               & \num{1111617}         & \num{1110692}      & 29.2       & 0.08          \\ \hline
TCDM Memory          & \num{889069}          & \num{889051}       & 23.3       & 0.00          \\ \hline
Cluster Interconnect & \num{160908}          & \num{158859}       & 4.2        & 1.29          \\ \hline
Cluster Peripherals  & \num{134202}          & \num{134353}       & 3.5        & -0.11         \\ \hline
DMA                  & \num{101388}          & \num{101522}       & 2.7        & -0.13         \\ \hline
Core Region          & \num{82600}           & \num{70805}        & 2.1        & 16.66         \\ \hline
Lockstep Unit        & \num{11668}           & -                  & 0.3        & -             \\
\hline
\\
\multicolumn{5}{l}{CORE LEVEL}
\\
\\ \hline
Core                 & \num{79970}           & \num{68335}        & 100        & 17.03         \\ \hline
IF Stage             & \num{6803}            & \num{6563}         & 8.5        & 3.66          \\ \hline
ID Stage             & \num{28358}           & \num{26129}        & 35.5       & 8.53          \\ \hline
EX Stage             & \num{37156}           & \num{29026}        & 46.5       & 28.01         \\ \hline
Load Store Unit      & \num{2104}            & \num{1930}         & 2.6        & 9.02          \\ \hline
CSR                  & \num{4959}            & \num{4629}         & 6.2        & 7.13\\
\hline
\end{tabular}

%% file: resnet_table.tex
\begin{tabular}{r|c|c|c|c}
\multicolumn{1}{c|}{\textbf{Processor}} &
  \textbf{Technology} &
  \textbf{ISA} &
  \textbf{Latency {[}ms{]}} &
  \multicolumn{1}{c}{\textbf{Energy {[}uJ{]}}} \\ \hline
\textit{\textbf{\begin{tabular}[c]{@{}r@{}}DUSTIN\\ (this work)\end{tabular}}} &
  \textit{\textbf{65 nm}} &
  \textit{\textbf{\begin{tabular}[c]{@{}c@{}}RISCV \\ + Custom ISA\end{tabular}}} &
  \textit{\textbf{3.80}} &
  \textit{\textbf{373}} \\ \hline
GAP9        & 22 nm & \begin{tabular}[c]{@{}c@{}}RISCV \\ + ACC.\end{tabular} & 0.62  & 40.4    \\ \hline
NDP9120-EVL & 40nm  & \begin{tabular}[c]{@{}c@{}}ARM \\ + ACC.\end{tabular}   & 5.12  & 139.4   \\ \hline
CORTEX M7   & 40nm  & ARM                                                     & 54.3  & 8707.3  \\ \hline
CORTEX M4   & 90nm  & ARM                                                     & 226.9 & 10681.6 \\ \hline
CORTEX M33  & 40nm  & ARM                                                     & 139.7 & 3642   
\end{tabular}

%% file: Soa_comparison_xpulpnn.tex
\begin{tabular}{c|c|c|cc|cc|cc}
 &
  \textbf{SleepRunner~\cite{SLEEPRUNNER}}&
  \textbf{SamurAI~\cite{SAMURAI}} &
  \multicolumn{2}{c|}{\textbf{VEGA~\cite{VEGA}}} &
  \multicolumn{2}{c|}{\textbf{XPULPNN~\cite{garofalo2020xpulpnn}}} &
  \multicolumn{2}{c}{\textbf{Dustin (this work)}} \\ \hline
\textbf{Technology} &
  \begin{tabular}[c]{@{}c@{}}CMOS 28nm \\ FD-SOI\end{tabular} &
  \begin{tabular}[c]{@{}c@{}}CMOS 28nm \\ FD-SOI\end{tabular} &
  \multicolumn{2}{c|}{\begin{tabular}[c]{@{}c@{}}CMOS 22nm \\ FD-SOI\end{tabular}} &
  \multicolumn{2}{c|}{\begin{tabular}[c]{@{}c@{}}CMOS 22nm \\ FD-SOI\end{tabular}} &
  \multicolumn{2}{c}{CMOS 65nm} \\ \hline
\textbf{Silicon Proven} &
  Yes &
  Yes &
  \multicolumn{2}{c|}{Yes} &
  \multicolumn{2}{c|}{No} &
  \multicolumn{2}{c}{Yes} \\ \hline
\textbf{Die Area} &
  0.68 mm${}^2$ &
  4.5 mm${}^2$ &
  \multicolumn{2}{c|}{12 mm${}^2$} &
  \multicolumn{2}{c|}{1.05 mm${}^2$} &
  \multicolumn{2}{c}{10 mm${}^2$} \\ \hline
\textbf{Apllications} &
  IoT GP &
  IoT GP + DNN &
  \multicolumn{2}{c|}{IoT GP + DNN} &
  \multicolumn{2}{c|}{IoT GP + DNN + QNN} &
  \multicolumn{2}{c}{IoT GP + DNN + QNN} \\ \hline
\textbf{CPU/ISA} &
  \begin{tabular}[c]{@{}c@{}}CM0DS \\ Thumb-2 subset\end{tabular} &
  \begin{tabular}[c]{@{}c@{}}1x RI5CY \\ RVC32IMFXpulp\end{tabular} &
  \multicolumn{2}{c|}{\begin{tabular}[c]{@{}c@{}}9x RI5CY \\ RVC32IMFXpulp\end{tabular}} &
  \multicolumn{2}{c|}{\begin{tabular}[c]{@{}c@{}}8x RI5CY\\ RV32ICMXpulpnn\end{tabular}} &
  \multicolumn{2}{c}{\begin{tabular}[c]{@{}c@{}}16x RI5CY \\ Mixed-Precision Extended\end{tabular}} \\ \hline
\textbf{Int Precision (bits)} &
  32 &
  8,16,32 &
  \multicolumn{2}{c|}{8,16,32} &
  \multicolumn{2}{c|}{2,4,8,16,32} &
  \multicolumn{2}{c}{\begin{tabular}[c]{@{}c@{}}2,4,8,16,32 \\ (plus mixed-precision)\end{tabular}} \\ \hline
\textbf{Supply Voltage} &
  0.4 - 0.8 V &
  0.45 - 0.9 V &
  \multicolumn{2}{c|}{0.5 - 0.8 V} &
  \multicolumn{2}{c|}{0.6 - 0.8 V} &
  \multicolumn{2}{c}{0.8 - 1.2 V} \\ \hline
\textbf{Max Frequency} &
  80 MHz &
  350 MHz &
  \multicolumn{2}{c|}{450 MHz} &
  \multicolumn{2}{c|}{400 MHz} &
  \multicolumn{2}{c}{60 - 205 MHz} \\ \hline
\textbf{Power Envelope} &
  320 uW &
  96 mW &
  \multicolumn{2}{c|}{49.4 mW} &
  \multicolumn{2}{c|}{19.3 - 41.6 mW} &
  \multicolumn{2}{c}{23 [14] - 156 [98] mW${}^3$} \\ \hline
\textbf{${}^1$Best Integer Performance} &
  31 MOPS(32b) &
  1.5 GOPS (8b)${}^2$ &
  \multicolumn{1}{c|}{15.6 GOPS (8b)} &
  \begin{tabular}[c]{@{}c@{}}5.61 GOPS (8x4/2b)\\ 3.12 GOPS (4x2b)\end{tabular} &
  \multicolumn{1}{c|}{\begin{tabular}[c]{@{}c@{}}23 GOPS (8b) \\ 43 GOPS (4b) \\ 72 GOPS (2b)\end{tabular}} &
  \begin{tabular}[c]{@{}c@{}}8.27 GOPS (8x4/2b)\\ 8.6 GOPS (4x2b)\end{tabular} &
  \multicolumn{1}{c|}{\textbf{\begin{tabular}[c]{@{}c@{}}15 GOPS (8b) \\ 30 GOPS (4b) \\ 58 GOPS (2b)\end{tabular}}} &
  \textbf{\begin{tabular}[c]{@{}c@{}}16.7 GOPS (8x4b)\\ 18.4 GOPS (8x2b)\\ 33.6 GOPS (4x2b)\end{tabular}} \\ \hline
\textbf{${}^1$Best Integer Efficiency} &
  \begin{tabular}[c]{@{}c@{}}97 MOPS/mW \\ @ 18.6 mops (32b)\end{tabular} &
  \begin{tabular}[c]{@{}c@{}}230 GOPS/W \\ @ 110 MOPS (8b)${}^2$\end{tabular} &
  \multicolumn{1}{c|}{\begin{tabular}[c]{@{}c@{}}614 GOPS/W \\ @ 7.6 GOPS (8b)\end{tabular}} &
  \begin{tabular}[c]{@{}c@{}}220 GOPS/W\\ @ 2,7 GOPS (8x4/2b)\\ 123 GOPS/W\\ @ 1.52 GOPS (4x2b)\end{tabular} &
  \multicolumn{1}{c|}{\begin{tabular}[c]{@{}c@{}}1111 GOPS/W (8b) \\ @ 11.4 GOPS \\ 2565 GOPS/W (4b)  \\ @ 21.7 GOPS \\ 3050 GOPS/W (2b)  \\ @ 36.2 GOPS\end{tabular}} &
  \begin{tabular}[c]{@{}c@{}}400 GOPS/W \\ @ 4.1 GOPS (8x4/2b)\\ 513 GOPS/W \\ @ 4.3 GOPS (4x2b)\end{tabular} &
  \multicolumn{1}{c|}{\textbf{\begin{tabular}[c]{@{}c@{}}303 GOPS/W (8b) @ \\ 4.4 GOPS \\ 570 GOPS/W (4b) @ \\ 8.8 GOPS \\ 1152 GOPS/W (2b) @ \\ 17.3 GOPS\end{tabular}}} &
  \textbf{\begin{tabular}[c]{@{}c@{}}345 GOPS/W \\ @ 5 GOPS (8x4b)\\ 379 GOPS/W\\ @ 5.5 GOPS (8x2b)\\ 640 GOPS/W\\ @ 10 GOPS (4x2b)\end{tabular}}
\end{tabular}

%% file: bio/bio.tex
\begin{IEEEbiography}[{\includegraphics[width=1in,height=1.25in,clip,keepaspectratio]{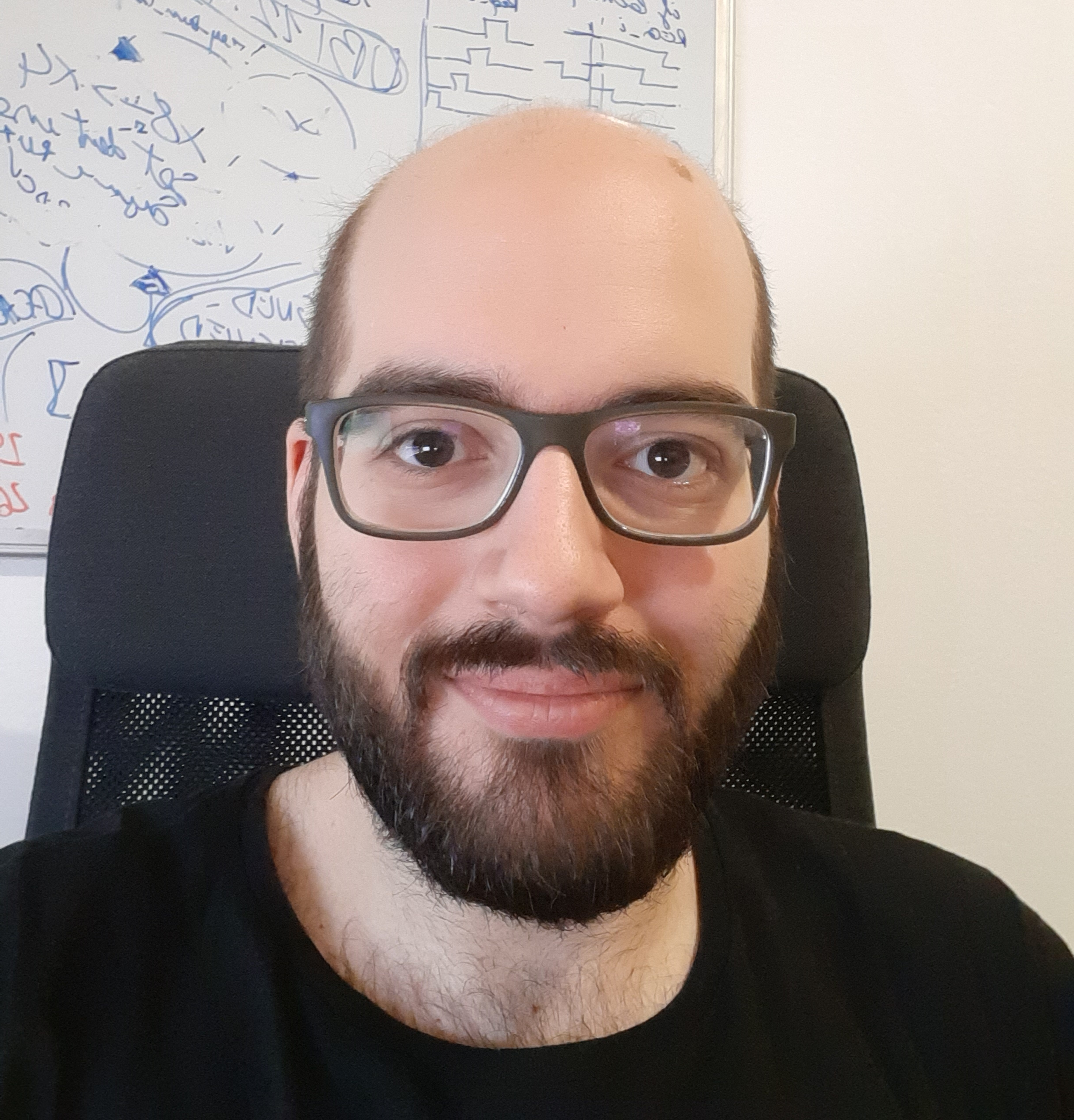}}]{Gianmarco Ottavi} recently started his Ph.D. in electronics Engineerig at the University of Bologna, Italy. After receiving his M.Sc. degree in 2019 he worked as a Research Fellow for two years in the Department of Electrical, Electronic and Information Engineering (DEI) in Bologna. Its research is focused on hardware design for efficient inference in low-power systems, where he developed specialized ISA extension for RISC-V and system-level implementation of In-memory computing accelerators.
\end{IEEEbiography}

\begin{IEEEbiography}[{\includegraphics[width=1in,height=1.25in,clip,keepaspectratio]{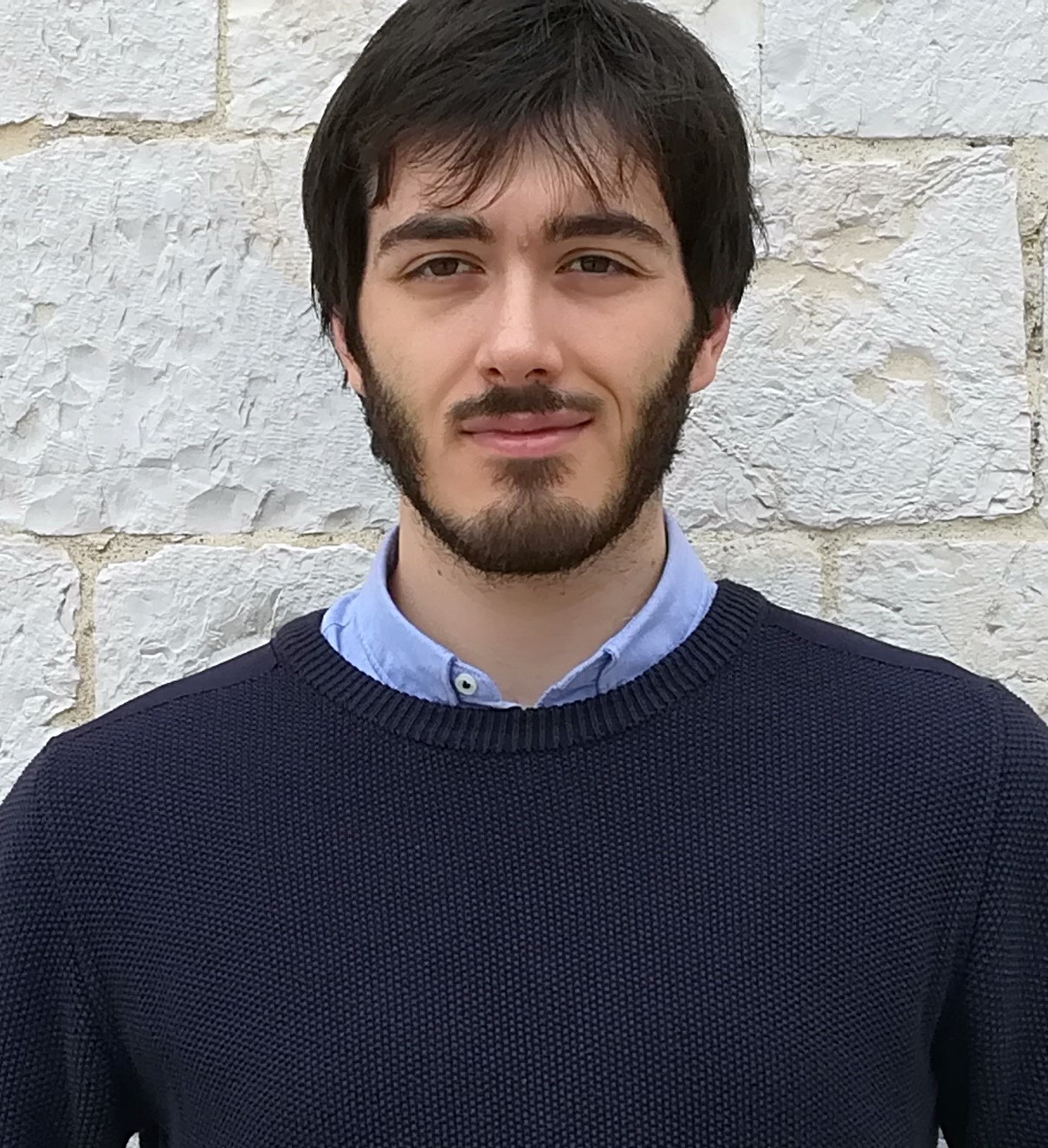}}]{Angelo Garofalo} received the B.Sc and M.Sc. degree in Electronic Engineering from the University of Bologna, Italy, in 2016 and 2018 respectively. He is currently working toward his Ph.D. degree at DEI, University of Bologna, Italy. His main research topic is Hardware-Software design of ultra-low-power multiprocessor systems on chip for edge AI. His research interests include Quantized Neural Networks, Hardware efficient Machine Learning, in-memory computing heterogeneous architectures and fully-programmable embedded architectures.
\end{IEEEbiography}
\vspace{-11mm}
\begin{IEEEbiography}[{\includegraphics[width=1in,height=1.25in,clip,keepaspectratio]{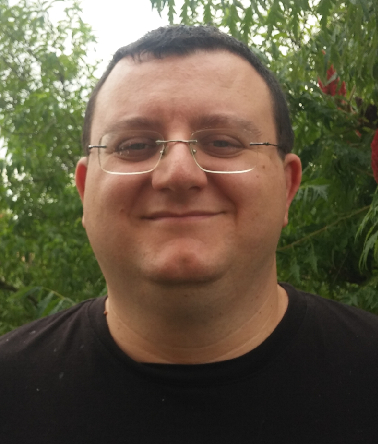}}]{Giuseppe Tagliavini} received a Ph.D. degree in Electronic Engineering from the University of Bologna, Bologna, Italy, in 2017. He is currently an Assistant Professor with the Department of Computer Science and Engineering (DISI) at the University of Bologna. He has co-authored over 40 papers in international conferences and journals. His main research interests include parallel programming models for embedded systems, run-time and compile-time optimization for multi/many-core platforms, HW/SW co-design of emerging computing architectures.
\end{IEEEbiography}
\vspace{-11mm}
\begin{IEEEbiography}[{\includegraphics[width=1in,height=1.25in,clip,keepaspectratio]{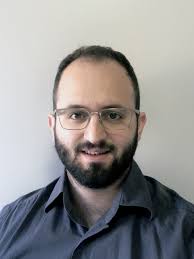}}]{Francesco Conti} received the Ph.D. degree in electronic engineering from the University of Bologna, Italy, in 2016. He is currently an Assistant Professor in the DEI Department of the University of Bologna. 
From 2016 to 2020, he held a research grant at the University of Bologna and a position as postdoctoral researcher at ETH Zurich.
His research focuses on the development of deep learning based intelligence on top of ultra-low power, ultra-energy efficient programmable SoCs. His research work has resulted in more than 40 publications in international conferences and journals and has been awarded several times, including the 2020 IEEE TCAS-I Darlington Best Paper Award.
\end{IEEEbiography}
\vspace{-11mm}
\begin{IEEEbiography}[{\includegraphics[width=1in,height=1.25in,clip,keepaspectratio]{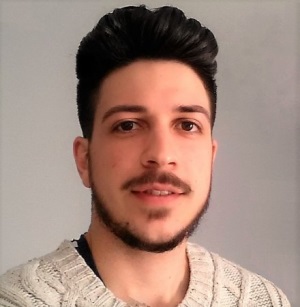}}]{Alfio Di Mauro} received the M.Sc.  degrees in Electronic Engineering from the Electronics and Telecommunications Department (DET) of Politecnico di Torino in 2016. Since September 2017, he is currently pursuing the Ph.D. at the Integrated System Laboratory (IIS) of the Swiss Federal Institute of Technology of Zurich. His research focuses on the design of digital Ultra-Low Power (ULP) System-on-Chip (SoC) for Event-Driven edge computing.
\end{IEEEbiography}
\vspace{-12mm}

\begin{IEEEbiography}[{\includegraphics[width=1in,height=1.25in,clip,keepaspectratio]{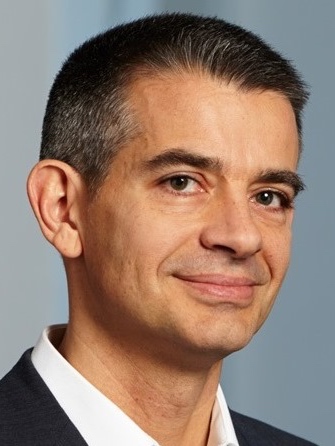}}]{Luca Benini} holds the chair of digital Circuits and systems at ETHZ and is Full Professor at the Università di Bologna. He received a PhD from Stanford University. Dr. Benini's research interests are in energy-efficient parallel computing systems, smart sensing micro-systems and machine learning hardware. He has published more than 1000 peer-reviewed papers and five books. He is a Fellow of the IEEE, of the ACM and a member of the Academia Europaea. He received the IEEE Mac Van Valkenburg award in 2016 and the ACM/IEEE A. Richard Newton Award in 2020.
\end{IEEEbiography}
\vspace{-9mm}
\begin{IEEEbiography}[{\includegraphics[width=1in,height=1.25in,clip,keepaspectratio]{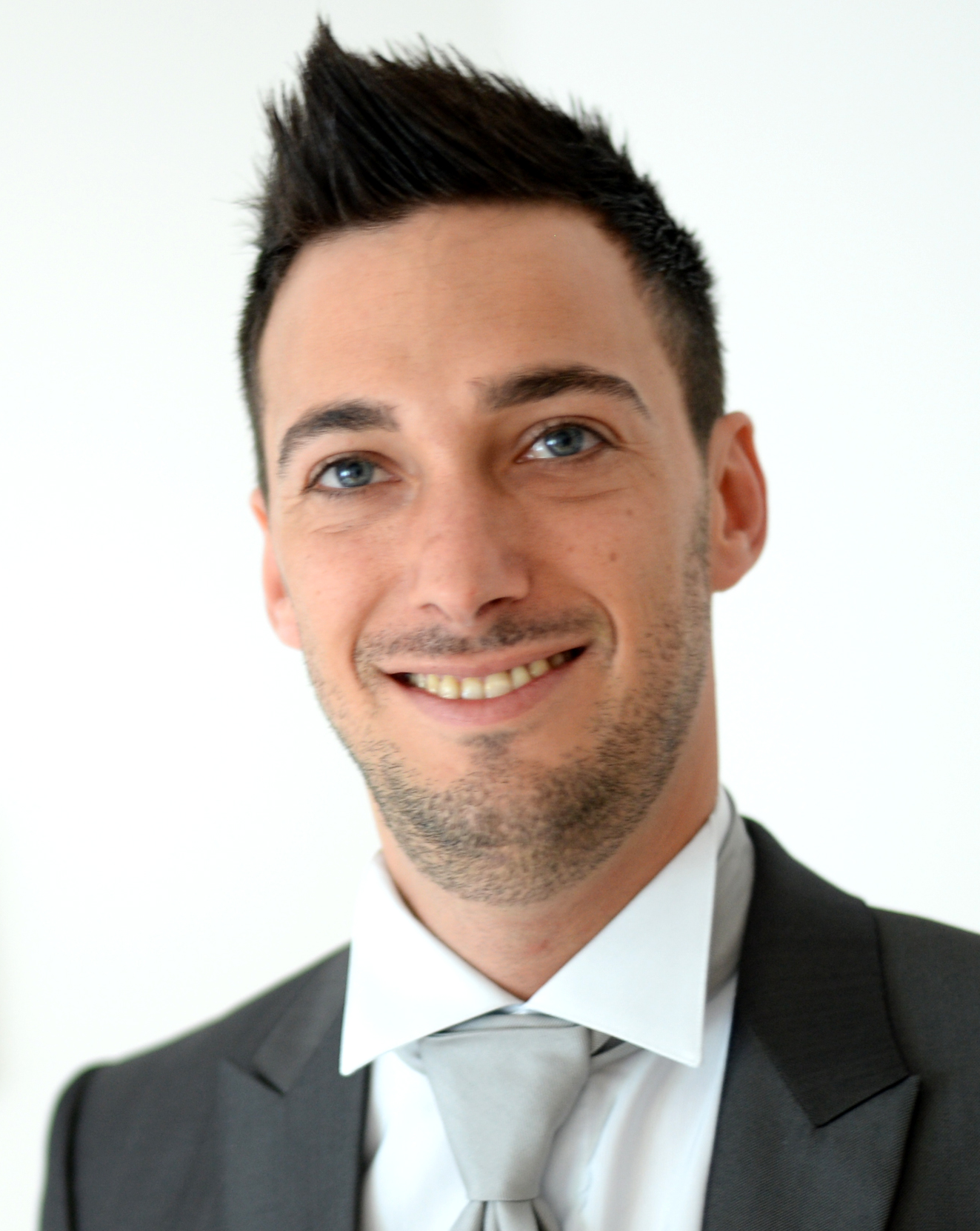}}]{Davide Rossi} received the Ph.D. degree from the University of Bologna, Bologna, Italy, in 2012. He has been a Post-Doctoral Researcher with the Department of Electrical, Electronic and Information Engineering “Guglielmo Marconi,” University of Bologna, since 2015, where he is currently an Assistant Professor. His research interests focus on energy-efficient digital architectures. In this field, he has published more than 100 papers in international peer-reviewed conferences and journals. He is recipient of Donald O. Pederson Best Paper Award 2018, 2020 IEEE TCAS Darlington Best Paper Award, 2020 IEEE TVLSI Prize Paper Award.
\end{IEEEbiography}

%% file: main.bbl
\begin{thebibliography}{10}
\providecommand{\url}[1]{#1}
\csname url@samestyle\endcsname
\providecommand{\newblock}{\relax}
\providecommand{\bibinfo}[2]{#2}
\providecommand{\BIBentrySTDinterwordspacing}{\spaceskip=0pt\relax}
\providecommand{\BIBentryALTinterwordstretchfactor}{4}
\providecommand{\BIBentryALTinterwordspacing}{\spaceskip=\fontdimen2\font plus
\BIBentryALTinterwordstretchfactor\fontdimen3\font minus
  \fontdimen4\font\relax}
\providecommand{\BIBforeignlanguage}[2]{{%
\expandafter\ifx\csname l@#1\endcsname\relax
\typeout{** WARNING: IEEEtran.bst: No hyphenation pattern has been}%
\typeout{** loaded for the language `#1'. Using the pattern for}%
\typeout{** the default language instead.}%
\else
\language=\csname l@#1\endcsname
\fi
#2}}
\providecommand{\BIBdecl}{\relax}
\BIBdecl

\bibitem{van2020bayesian}
M.~van Baalen, C.~Louizos, M.~Nagel, R.~A. Amjad, Y.~Wang, T.~Blankevoort, and
  M.~Welling, ``{Bayesian bits: Unifying quantization and pruning},''
  \emph{arXiv preprint arXiv:2005.07093}, 2020.

\bibitem{cai2020rethinking}
Z.~Cai and N.~Vasconcelos, ``{Rethinking differentiable search for
  mixed-precision neural networks},'' in \emph{Proceedings of the IEEE/CVF
  Conference on Computer Vision and Pattern Recognition}, 2020, pp. 2349--2358.

\bibitem{GANPU}
S.~{Kang}, D.~{Han}, J.~{Lee}, D.~{Im}, S.~{Kim}, S.~{Kim}, and H.~{Yoo},
  ``{{7.4 GANPU: A 135TFLOPS/W Multi-DNN Training Processor for GANs with
  Speculative Dual-Sparsity Exploitation}},'' in \emph{2020 IEEE International
  Solid- State Circuits Conference - (ISSCC)}, 2020, pp. 140--142.

\bibitem{lee2018unpu}
J.~Lee, C.~Kim, S.~Kang, D.~Shin, S.~Kim, and H.-J. Yoo, ``{UNPU: A 50.6 TOPS/W
  unified deep neural network accelerator with 1b-to-16b fully-variable weight
  bit-precision},'' in \emph{2018 IEEE International Solid-State Circuits
  Conference-(ISSCC)}.\hskip 1em plus 0.5em minus 0.4em\relax IEEE, 2018, pp.
  218--220.

\bibitem{garofalo2020pulp}
A.~Garofalo, M.~Rusci, F.~Conti, D.~Rossi, and L.~Benini, ``{PULP-NN:
  accelerating quantized neural networks on parallel ultra-low-power RISC-V
  processors},'' \emph{Philosophical Transactions of the Royal Society A}, vol.
  378, no. 2164, p. 20190155, 2020.

\bibitem{lai2018cmsis}
L.~Lai, N.~Suda, and V.~Chandra, ``{CMSIS-NN: Efficient Neural Network Kernels
  for Arm Cortex-M CPUs},'' \emph{arXiv preprint arXiv:1801.06601}, 2018.

\bibitem{ARMHELIUM}
ARM, ``{ARM Helium},'' \url{https://www.arm.com/technologies/helium}. Last
  accessed on Sept. 22.

\bibitem{garofalo2020xpulpnn}
A.~Garofalo, G.~Tagliavini, F.~Conti, L.~Benini, and D.~Rossi, ``{XpulpNN:
  Enabling Energy Efficient and Flexible Inference of Quantized Neural Networks
  on RISC-V Based IoT End Nodes},'' \emph{IEEE Transactions on Emerging Topics
  in Computing}, vol.~9, no.~3, pp. 1489--1505, 2021.

\bibitem{bruschi2020enabling}
N.~Bruschi, A.~Garofalo, F.~Conti, G.~Tagliavini, and D.~Rossi, ``{Enabling
  mixed-precision quantized neural networks in extreme-edge devices},'' in
  \emph{Proceedings of the 17th ACM International Conference on Computing
  Frontiers}, 2020, pp. 217--220.

\bibitem{howard2017mobilenets}
A.~G. Howard, M.~Zhu, B.~Chen, D.~Kalenichenko, W.~Wang, T.~Weyand,
  M.~Andreetto, and H.~Adam, ``Mobilenets: Efficient convolutional neural
  networks for mobile vision applications,'' \emph{arXiv preprint
  arXiv:1704.04861}, 2017.

\bibitem{sandler2018mobilenetv2}
M.~Sandler, A.~Howard, M.~Zhu, A.~Zhmoginov, and L.-C. Chen, ``Mobilenetv2:
  Inverted residuals and linear bottlenecks,'' in \emph{Proceedings of the IEEE
  conference on computer vision and pattern recognition}, 2018, pp. 4510--4520.

\bibitem{wang2019haq}
K.~Wang, Z.~Liu, Y.~Lin, J.~Lin, and S.~Han, ``{Haq: Hardware-aware automated
  quantization with mixed precision},'' in \emph{Proceedings of the IEEE
  conference on computer vision and pattern recognition}, 2019, pp. 8612--8620.

\bibitem{wu2018mixed}
B.~Wu, Y.~Wang, P.~Zhang, Y.~Tian, P.~Vajda, and K.~Keutzer, ``Mixed precision
  quantization of convnets via differentiable neural architecture search,''
  \emph{arXiv preprint arXiv:1812.00090}, 2018.

\bibitem{naumov2018periodic}
M.~Naumov, U.~Diril, J.~Park, B.~Ray, J.~Jablonski, and A.~Tulloch, ``On
  periodic functions as regularizers for quantization of neural networks,''
  \emph{arXiv preprint arXiv:1811.09862}, 2018.

\bibitem{dong2019hawq}
Z.~Dong, Z.~Yao, A.~Gholami, M.~W. Mahoney, and K.~Keutzer, ``Hawq: Hessian
  aware quantization of neural networks with mixed-precision,'' in
  \emph{Proceedings of the IEEE/CVF International Conference on Computer
  Vision}, 2019, pp. 293--302.

\bibitem{dong2020hawq}
Z.~Dong, Z.~Yao, D.~Arfeen, A.~Gholami, M.~W. Mahoney, and K.~Keutzer,
  ``Hawq-v2: Hessian aware trace-weighted quantization of neural networks,''
  \emph{Advances in neural information processing systems}, vol.~33, pp.
  18\,518--18\,529, 2020.

\bibitem{yao2021hawq}
Z.~Yao, Z.~Dong, Z.~Zheng, A.~Gholami, J.~Yu, E.~Tan, L.~Wang, Q.~Huang,
  Y.~Wang, M.~Mahoney \emph{et~al.}, ``Hawq-v3: Dyadic neural network
  quantization,'' in \emph{International Conference on Machine Learning}.\hskip
  1em plus 0.5em minus 0.4em\relax PMLR, 2021, pp. 11\,875--11\,886.

\bibitem{choi2018bridging}
J.~Choi, P.~I.-J. Chuang, Z.~Wang, S.~Venkataramani, V.~Srinivasan, and
  K.~Gopalakrishnan, ``Bridging the accuracy gap for 2-bit quantized neural
  networks (qnn),'' \emph{arXiv preprint arXiv:1807.06964}, 2018.

\bibitem{moons201714}
B.~Moons, R.~Uytterhoeven, W.~Dehaene, and M.~Verhelst, ``{14.5 Envision: A
  0.26-to-10TOPS/W subword-parallel dynamic-voltage-accuracy-frequency-scalable
  Convolutional Neural Network processor in 28nm FDSOI},'' in \emph{2017 IEEE
  International Solid-State Circuits Conference (ISSCC)}.\hskip 1em plus 0.5em
  minus 0.4em\relax IEEE, 2017, pp. 246--247.

\bibitem{yin20171}
S.~Yin, P.~Ouyang, S.~Tang, F.~Tu, X.~Li, L.~Liu, and S.~Wei, ``{A 1.06-to-5.09
  TOPS/W reconfigurable hybrid-neural-network processor for deep learning
  applications},'' in \emph{2017 Symposium on VLSI Circuits}.\hskip 1em plus
  0.5em minus 0.4em\relax IEEE, 2017, pp. C26--C27.

\bibitem{sharify2018loom}
S.~Sharify, A.~D. Lascorz, K.~Siu, P.~Judd, and A.~Moshovos, ``{Loom:
  Exploiting weight and activation precisions to accelerate convolutional
  neural networks},'' in \emph{2018 55th ACM/ESDA/IEEE Design Automation
  Conference (DAC)}.\hskip 1em plus 0.5em minus 0.4em\relax IEEE, 2018, pp.
  1--6.

\bibitem{chen2019eyeriss}
Y.-H. Chen, T.-J. Yang, J.~Emer, and V.~Sze, ``{Eyeriss v2: A flexible
  accelerator for emerging deep neural networks on mobile devices},''
  \emph{IEEE Journal on Emerging and Selected Topics in Circuits and Systems},
  vol.~9, no.~2, pp. 292--308, 2019.

\bibitem{LatticeSENSEAI}
Lattice, ``{Lattice sensAI Delivers 10X Performance Boost for Low-Power, Smart
  IoT Devices at the Edge},''
  \url{https://www.latticesemi.com/About/Newsroom/PressReleases/2019/201911sensAI}.
  Last accessed on Sept. 20.

\bibitem{VEGA}
D.~Rossi, F.~Conti, M.~Eggimann, A.~Di~Mauro, G.~Tagliavini, S.~Mach,
  M.~Guermandi, A.~Pullini, I.~Loi, J.~Chen \emph{et~al.}, ``{Vega: A Ten-Core
  SoC for IoT Endnodes With DNN Acceleration and Cognitive Wake-Up From
  MRAM-Based State-Retentive Sleep Mode},'' \emph{IEEE Journal of Solid-State
  Circuits}, 2021.

\bibitem{dogan2013synchronizing}
A.~Y. Dogan, R.~Braojos, J.~Constantin, G.~Ansaloni, A.~Burg, and D.~Atienza,
  ``{Synchronizing code execution on ultra-low-power embedded multi-channel
  signal analysis platforms},'' in \emph{2013 Design, Automation \& Test in
  Europe Conference \& Exhibition (DATE)}.\hskip 1em plus 0.5em minus
  0.4em\relax IEEE, 2013, pp. 396--399.

\bibitem{cavalcante2021mempool}
M.~Cavalcante, S.~Riedel, A.~Pullini, and L.~Benini, ``Mempool: A shared-l1
  memory many-core cluster with a low-latency interconnect,'' in \emph{2021
  Design, Automation \& Test in Europe Conference \& Exhibition (DATE)}.\hskip
  1em plus 0.5em minus 0.4em\relax IEEE, 2021, pp. 701--706.

\bibitem{benini2012p2012}
L.~Benini, E.~Flamand, D.~Fuin, and D.~Melpignano, ``P2012: Building an
  ecosystem for a scalable, modular and high-efficiency embedded computing
  accelerator,'' in \emph{2012 Design, Automation \& Test in Europe Conference
  \& Exhibition (DATE)}.\hskip 1em plus 0.5em minus 0.4em\relax IEEE, 2012, pp.
  983--987.

\bibitem{meinerzhagen2010towards}
P.~Meinerzhagen, C.~Roth, and A.~Burg, ``{Towards generic low-power
  area-efficient standard cell based memory architectures},'' in \emph{2010
  53rd IEEE International Midwest Symposium on Circuits and Systems}.\hskip 1em
  plus 0.5em minus 0.4em\relax IEEE, 2010, pp. 129--132.

\bibitem{glaser2020energy}
F.~Glaser, G.~Tagliavini, D.~Rossi, G.~Haugou, Q.~Huang, and L.~Benini,
  ``{Energy-Efficient Hardware-Accelerated Synchronization for Shared-L1-Memory
  Multiprocessor Clusters},'' \emph{IEEE Transactions on Parallel and
  Distributed Systems}, vol.~32, no.~3, pp. 633--648, 2020.

\bibitem{banbury2021mlperf}
C.~Banbury, V.~J. Reddi, P.~Torelli, J.~Holleman, N.~Jeffries, C.~Kiraly,
  P.~Montino, D.~Kanter, S.~Ahmed, D.~Pau \emph{et~al.}, ``Mlperf tiny
  benchmark,'' \emph{arXiv preprint arXiv:2106.07597}, 2021.

\bibitem{capotondi2020cmix}
A.~Capotondi, M.~Rusci, M.~Fariselli, and L.~Benini, ``Cmix-nn: Mixed
  low-precision cnn library for memory-constrained edge devices,'' \emph{IEEE
  Transactions on Circuits and Systems II: Express Briefs}, vol.~67, no.~5, pp.
  871--875, 2020.

\bibitem{SLEEPRUNNER}
D.~{Bol}, M.~{Schramme}, L.~{Moreau}, P.~{Xu}, R.~{Dekimpe}, R.~{Saeidi},
  T.~{Haine}, C.~{Frenkel}, and D.~{Flandre}, ``{{SleepRunner: A 28-nm FDSOI
  ULP Cortex-M0 MCU With ULL SRAM and UFBR PVT Compensation for
  2.6-3.6-$\mu$W/DMIPS 40-80-MHz Active Mode and 131-nW/kB Fully Retentive
  Deep-Sleep Mode}},'' \emph{IEEE Journal of Solid-State Circuits}, pp. 1--1,
  2021.

\bibitem{SAMURAI}
I.~{Miro-Panades} \emph{et~al.}, ``{{SamurAI: A 1.7MOPS-36GOPS Adaptive
  Versatile IoT Node with 15,000× Peak-to-Idle Power Reduction, 207ns Wake-Up
  Time and 1.3TOPS/W ML Efficiency}},'' in \emph{2020 IEEE Symposium on VLSI
  Circuits}, 2020, pp. 1--2.

\bibitem{WOLF}
A.~{Pullini}, D.~{Rossi}, I.~{Loi}, G.~{Tagliavini}, and L.~{Benini},
  ``{{Mr.Wolf: An Energy-Precision Scalable Parallel Ultra Low Power SoC for
  IoT Edge Processing}},'' \emph{IEEE Journal of Solid-State Circuits},
  vol.~54, no.~7, pp. 1970--1981, 2019.

\end{thebibliography}
